\documentclass[10pt,twocolumn,letterpaper]{article}

\usepackage{cvpr}              
\usepackage{multirow}

\usepackage{mathtools}
\usepackage{amsfonts}
\usepackage{bm}

\usepackage[inline]{enumitem}
\usepackage{xcolor}

\usepackage{xspace}

\usepackage{wrapfig}

\usepackage{multirow, makecell}

\usepackage{algorithm}
\usepackage{algpseudocode}

\usepackage[accsupp]{axessibility}

\def\Reals{\mathbb{R}}
\newcommand{\myvec}[1]{\bm{#1}}
\newcommand{\mymat}[1]{\mathbf{#1}}
\newcommand{\mesh}[1]{\mathcal{#1}}
\def\M{\mesh{M}}
\def\N{\mesh{N}}
\def\FM{\mathcal{F}(\M)}
\def\FN{\mathcal{F}(\N)}
\newcommand{\evecs}[1]{\Phi_{#1}}
\newcommand{\evals}[1]{\Lambda_{#1}}
\def\pinv{^{\dagger}}

\def\eg{\textit{e.g.}\xspace}
\def\ie{\textit{i.e.}\xspace}
\def\etal{\textit{et al.}\xspace}

\definecolor{cvprblue}{rgb}{0.21,0.49,0.74}
\usepackage[pagebackref,breaklinks,colorlinks,allcolors=cvprblue]{hyperref}

\title{Volumetric Functional Maps}

\author{Filippo Maggioli\\
Pegaso University\\
Naples, Italy\\
{\tt\small maggioli.filippo@gmail.com}
\and
Simone Melzi\\
University of Milano-Bicocca\\
Milan, Italy\\
{\tt\small simone.melzi@unimib.it}
\and
Marco Livesu\\
CNR IMATI\\
Genoa, Italy\\
{\tt\small marco.livesu@gmail.com}
}

\begin{document}
\maketitle
\begin{abstract}
Computing volumetric correspondences between 3D shapes is a prominent tool for medical and industrial applications. In this work, we pave the way for spectral volume mapping, extending for the first time the surface-based functional maps framework. We show that the eigenfunctions of the volumetric Laplace operator define a functional space that is suitable for high-quality signal transfer. We also experiment with various techniques that edit this functional space, porting them to volume domains. We validate our method on novel volumetric datasets and on tetrahedralizations of well established surface datasets, also showcasing practical applications involving both discrete and continuous signal mapping, for segmentation transfer, mesh connectivity transfer and solid texturing. Finally, we show that the volumetric spectrum greatly improves the accuracy for classical shape matching tasks among surfaces, consistently outperforming surface-only spectral methods.
\end{abstract}

\begin{figure}[t]
\centering
\includegraphics[width=0.95\columnwidth]{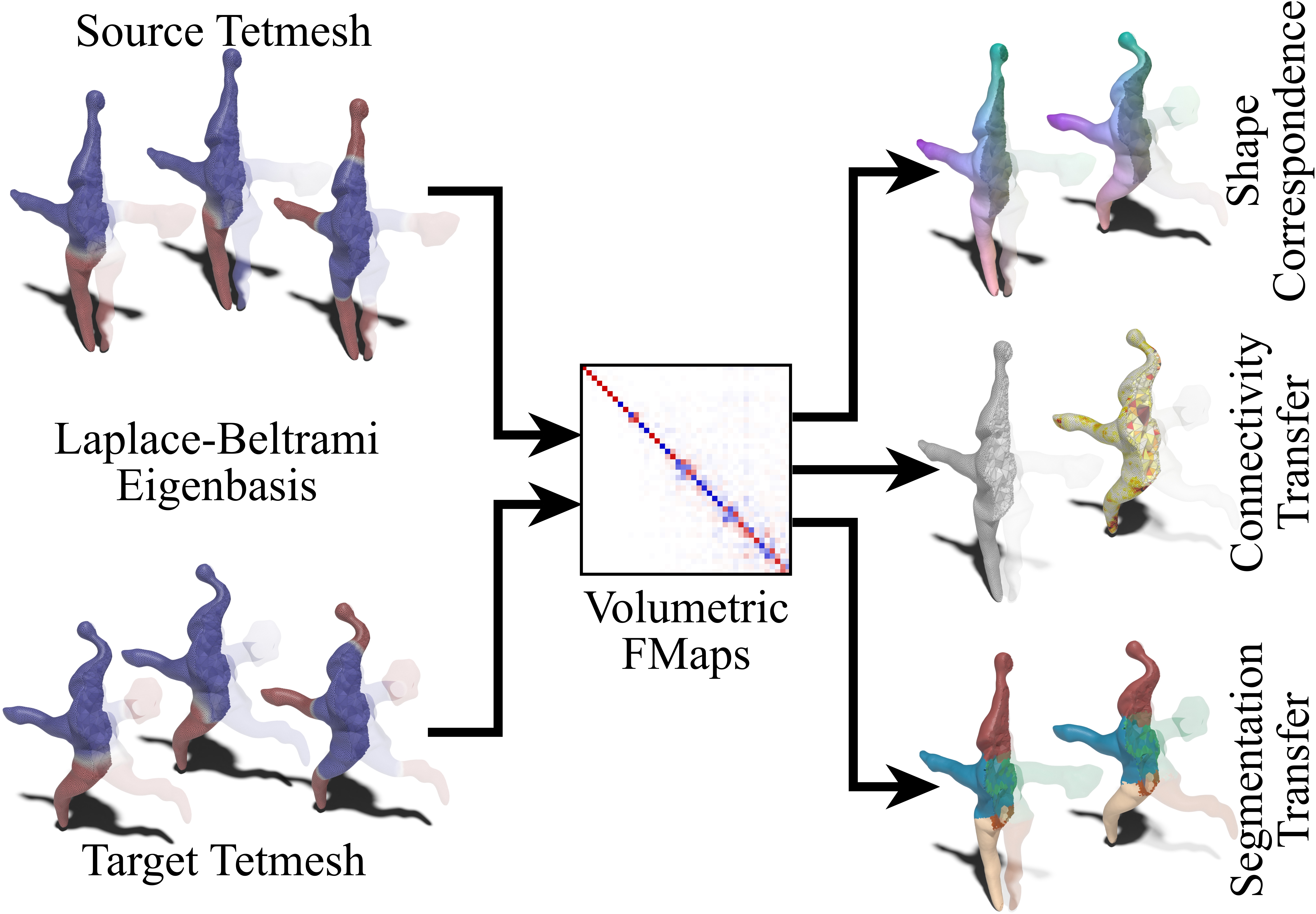}
\caption{Visual representation of our pipeline. The eigenfunctions of the LBO for volume meshes (left) are used to compute a volumetric functional map (middle). Basis alignment is exploited for several tasks: volumetric correspondences, piece-wise linear maps, and volumetric segmentation transfer (right).}
\label{fig:teaser}
\end{figure}

\section{Introduction}\label{sec:intro}

The computation of volumetric correspondences plays a central role for many applications.
Volume mappings enable non-invasive medical diagnosis, by warping the digital copy of a human organ into its canonical position without extracting it from the patient's body~\cite{abulnaga2021volumetric}. They are also important for meshing~\cite{pietroni2022hex}, statistical shape analysis~\cite{huang2019limit} and Industry 4.0, where physical simulations computed on predictive digital twins must be aligned with other digital views for optimized decision making and failure avoidance~\cite{rasheed2020digital}.



Maps between 3D objects are often represented either by exploiting one-to-one per vertex correspondences between two simplicial meshes with same connectivity, defining a so called piece-wise map (\cref{ssec:pl_maps}), 
or putting into correspondence two sets of basis functions to express the same signal in two alternative domains, as done by spectral approaches for surface meshes~\cite{ovsjanikov2012functional}.
Both piece-wise and spectral approaches have been extensively studied in the surface setting and can nowadays be considered rather mature, as proved by numerous academic and commercial tools that incorporate them. 
Conversely, the equivalent task of computing volumetric correspondences 
largely remains a scientific open problem~\cite[\S8.2]{pietroni2022hex}. It is worth noticing that the LDDMM framework has been successfully applied to match dense volumetric data (\eg, MRI/CRT scans), although diffeomorphic maps are limited to domains with grid-like structures and hardly generalize to more flexible representations like tetrahedral meshes\cite{ceritoglu2013computational,stouffer2022projective,piyadasa2025constrained,ceritoglu2010large}.

In recent years many attempts have been made to extend piece-wise linear approaches to volumes. This task has proved to be extremely challenging and no fully robust and efficient methods have been devised yet (\cref{ssec:pl_maps}).
To our knowledge no attempt has been made to extend functional mapping to volumes yet (\cref{ssec:func_maps}). 
We propose it for the first time and show that, conversely from the piece-wise linear setting, functional mapping naturally scales to volumes, providing a theoretically sound and practically useful platform for the computation of voulmetric correspondences.

Using the spectrum of the volumetric Laplacian, we define a functional space that is completely agnostic of the underlying discrete mesh, and we use it to create correspondences between solid meshes with different density and connectivity. We also show that many recent developments used to improve the functional maps framework~\cite{melzi2019zoomout,maggioli2021orthogonalized} can be exploited on volumes too, allowing to find an optimal balance between map quality and performance. 

We validate our volumetric pipeline (\cref{fig:teaser}) on a variety of datasets, including tetrahedralizations of prominent reference benchmarks widely used for shape correspondence tasks. Besides assessing the accuracy and performances of our tool on the ground truth (\cref{sec:results}), we showcase various practical applications, including connectivity transfer of a given volume mesh to a target domain (\cref{fig:volmap-cactus}), segmentation transfer from a statistical medical shape to acquired organs (\cref{fig:brains}) and solid texturing (\cref{fig:newton}).

All in all, this article opens for the first time to the computation of volumetric correspondences with spectral approaches. To this end, we believe our contributions will foster new research in this direction, positioning volume spectral mapping as a flexible and practical alternative to existing piece-wise linear approaches. To facilitate this endeavor, we release our reference code at \emph{anonymous}.

\section{Related Works}\label{sec:related}
We list the main approaches to compute piece-wise maps and functional maps highlighting, for both classes of methods, existing attempts (or lack thereof) to scale to volumes.

\subsection{Piece-wise Maps}
\label{ssec:pl_maps}
Piece-wise maps define correspondences between two meshes by starting from explicit per vertex correspondences and extending the map inside mesh elements by means of barycentric interpolation. This operation is uniquely defined for simplices in any dimension~\cite{hormann2014barycentric}, therefore it works for both surface (triangle) and volume (tetrahedral) meshes.

A variety of algorithms can robustly generate piece-wise surface maps to elementary domains such as triangles~\cite{finnendahl2023efficient}, convex polygons~\cite{tutte1963draw,floater1997parametrization,livesu2024stripe,shen2019progressive}, star-shaped polygons~\cite{Liv24} and spheres~\cite{gotsman2003fundamentals,praun2003spherical}. Maps between general shapes are then obtained via composition, using these elementary shapes as intermediate domains~\cite{schmidt2019distortion,schreiner2004inter,weber2014locally,lee1998maps,khodakovsky2003globally,praun2001consistent,schmidt2020inter,Liv20,garner2005topology,shi2016hyperbolic,aigerman2016hyperbolic,li2008globally,schmidt2023surface,kraevoy2004cross}. 

Lifting this idea to volumes is possible in principle but extremely challenging in practice. Even constructing an injective map to a convex polyhedron has proved to be surprisingly difficult~\cite{Liv20b,CL23,nigolian2024progressive}. Surface methods such as the Tutte embedding are known to fail on general volume meshes~\cite{alexa2023tutte,chilakamarri1995three}.
To date, the only robust volume embedding algorithms are~\cite{nigolian2023expansion,nigolian2024progressive,hinderink2023galaxy,campen2016bijective}, but all of them are unsuitable for real applications, due the massive use of exact numerical constructions and exponential mesh refinement~\cite{meloni2024extent}. 

A less robust but more practical alternative consists in initializing a non injective map directly with a fast approximation algorithm, and then relocate the vertices to fix flipped or vanishing elements. 
This problem, 
often referred to as mesh \emph{untangling}~\cite{knupp2001hexahedral,pietroni2022hex}, is ill-posed in general, as 
valid solutions may not exist for a fixed mesh connectivity~\cite{livesu2020mapping}. 
Existing volume approaches operate on convex subsets of the feasible region~\cite{kovalsky2014controlling}, minimize flip-preventing energies~\cite{du2020lifting,garanzha2021foldover,abulnaga2023symmetric}, or project onto the space of bounded distortion mappings~\cite{aigerman2013injective}. Due to the problem's non linearity and the absence of a feasible starting solution, none of these methods guarantees the correctness of the result. Multiple failure cases have been reported by testing the most prominent existing algorithms on available datasets~\cite[\S6.3]{nigolian2023expansion}. In \cref{ssec:conn_trans} we show that bootstrapping this pipeline with our functional mapping dramatically increases the success rate of state of the art volume untangling algorithms.

\subsection{Functional Maps}
\label{ssec:func_maps}
Given two discrete surfaces, shape matching aims to estimate a semantic point-wise correspondence between them. The more general (and challenging) scenario occurs when the two shapes are related by a non-rigid deformation.
Many applications are related to this task, from statistical shape analysis~\cite{Bogo:CVPR:2014} and medical imaging~\cite{magnet2023assessing} to deformation transfer~\cite{sumner04deformation,sundararaman24deformation}, among many others~\cite{sahilliouglu2020recent,deng2022survey}.
Among different approaches, functional mapping~\cite{ovsjanikov2012functional,ovsjanikov:2016:fmapsmatching} has attracted significant attention.
This framework aims to solve the shape matching problem by estimating a correspondence between functions defined on the surfaces, rather than point-wise correspondences, which are hard to optimize for. Exploiting a basis for the space of functions, for which a standard choice arises from the extension of Fourier theory to non-Euclidean domains~\cite{Taubin95signal,Levy10spectral}, the correspondence among shapes can be compactly encoded in a matrix with dimensions equal to the basis size. 
Following this direction, a variety of works try to improve the framework by defining additional functional constraints~\cite{nogneng2017informative,ren:2018:continuousfmaps,donati:2022:complexfmaps}, proposing alternative bases~\cite{melzi:2018:localeigs,nogneng2018improved,maggioli2021orthogonalized,marin2019cmh,marin2020correspondence}, or defining alternative procedures to extract point correspondences from a given functional map~\cite{rodola2017regularized,melzi2019zoomout,ren:2020:maptree,ren:2021:discreteopt,ezuz:2017:denoisemaps,pai2021sikhorn}. Recently, two alternatives for scaling the functional maps approach to high-resolution meshes~\cite{magnet:2023:scalablefmaps,maggioli2025rematching}  addressed the scalability issues of functional mapping. 
The functional maps framework additionally gives rise to a family of data-driven approaches that exploit machine learning techniques to solve the shape matching problem exploiting the functional representation as a prior~\cite{litany2017deep,donati:2020:deepfmaps,sharp2022diffusionnet,cao2024spectral,caobernerd23unsup,attaiki2024shape}. Despite its success, to the best of our knowledge, this framework has never been applied to volumes, leaving its potential impact fully unexplored in volume data processing. 

\section{Background and Notation}\label{sec:background}

We discretize volumes as tetrahedral meshes $\M = (V_{\M}, T_{\M})$, where:
\begin{itemize*}[label={}]
    \item $V_{\M} \subset \Reals^3$ is a set of vertices in 3D space;
    \item $T_{\M} \subset V_{\M}^4$ is a set of tetrahedra connecting the vertices in $V_{\M}$.
\end{itemize*}

\begingroup
\setlength{\intextsep}{-0.8pt}%
\setlength{\columnsep}{0pt}%
\begin{wrapfigure}[5]{r}[0pt]{0.43\columnwidth}
    \centering
    \vspace{-3em}%
    \includegraphics[width=0.4\columnwidth]{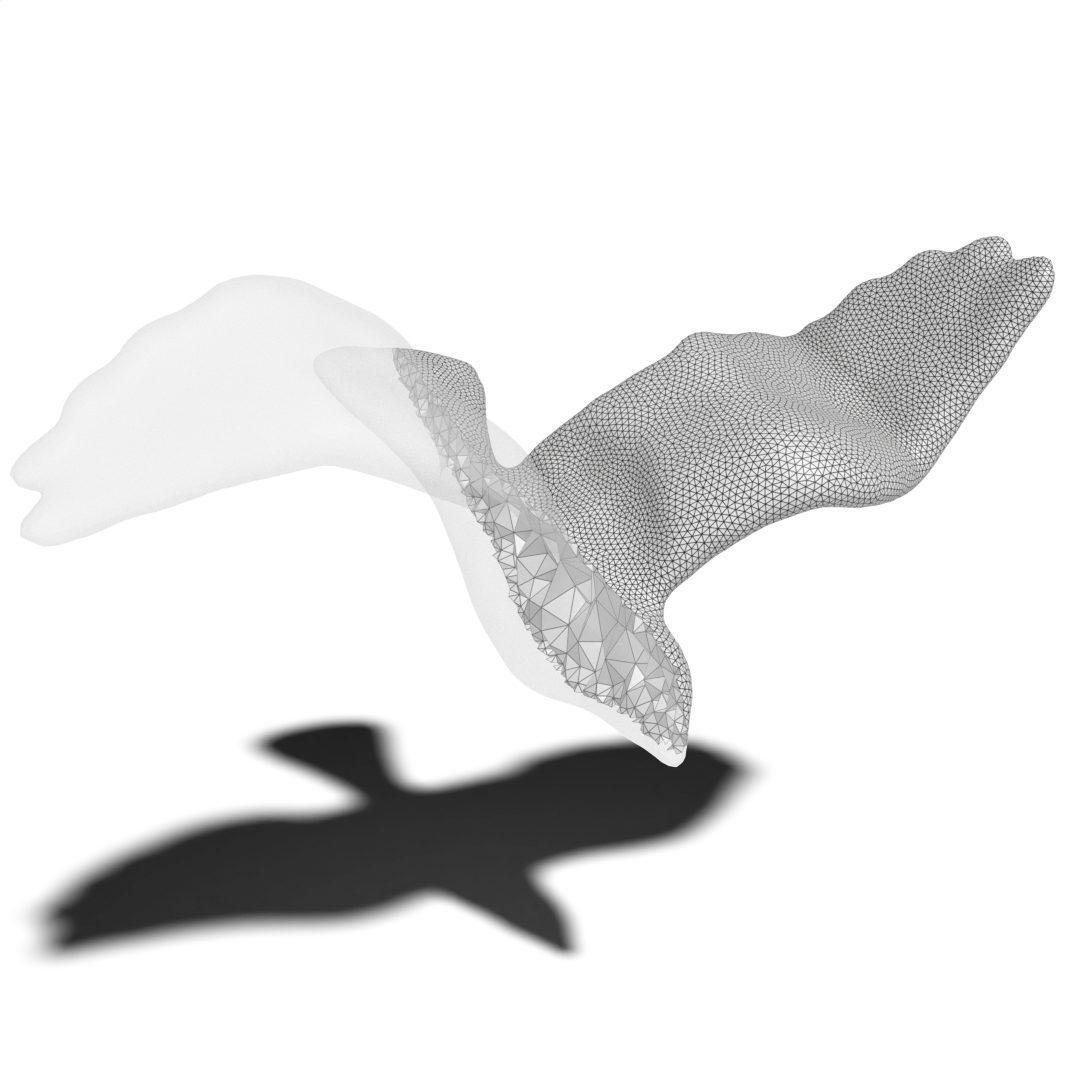}
\end{wrapfigure}
We refer to the external surface of $\M$ as the triangular mesh $\partial\M = (V_{\partial\M}, F_{\partial\M})$, where
\begin{itemize*}[label={}]
    \item $V_{\partial\M} \subset V_{\M}$ is the set of surface vertices of $\M$;
    \item $F_{\partial\M} \subset V_{\partial\M}^3$ is the set of surface triangles defined by the tetrahedra in $T_{\M}$.
\end{itemize*}

The inset figure depicts a section of a tetrahedral mesh, with the original external surface shown in transparency.
\endgroup


A scalar function $f : V_{\M} \to \Reals$ over a volume mesh $\M$ is discretized as a vector $\myvec{f} \in \Reals^{|V_{\M}|}$, and its restriction $f_{\partial} : V_{\partial\M} \to \Reals$ to the external surface $\partial\M$ is discretized as a vector $\myvec{f}_{\partial} \in \Reals^{|V_{\partial\M}|}$.
Given a tetrahedral (or triangular) mesh $\M$, and given two vertices $v_i, v_j \in \M$, we refer to the geodesic distance between $v_i$ and $v_j$ over $\M$ as $d_{\M}(v_i, v_j)$. There exist several algorithms to estimate such a distance on discrete meshes~\cite{crane2020survey}. For the sake of simplicity, in the remainder of the paper we will always approximate $d_{\M}(v_i, v_j)$ using Dijkstra's algoritm~\cite{dijkstra:1959:shortestpath} on the edge connectivity of $\M$.

\paragraph*{Shape Matching and Correspondences.}
Given two surface meshes $\M = (V_{\M}, F_{\M})$ and $\N = (V_{\N}, F_{\N})$, the problem of \emph{shape matching} is the problem of finding a \emph{correspondence} $\pi : V_{\M} \to V_{\N}$ that maps vertices of $\M$ into vertices of $\N$~\cite{loncaric1998survey}. In most cases, the correspondence $\pi$ has to satisfy semantic constraints or has to be an \emph{isometry}~\cite{sahilliouglu2020recent,deng2022survey}, that is a correspondence $\pi$ preserving the geodesic distances between pairs of vertices. Namely
\begin{equation}
    \forall v_i, v_j \in V_{\M},\quad
    d_{\M}(v_i, v_j)
    =
    d_{\N}(\pi(v_i), \pi(v_j))\,.
\end{equation}

\paragraph*{Functional Maps.}
The problem of finding an isometric correspondence, in its general setting, is NP-hard~\cite{benkner2021q}. Many solutions have been developed for computing an approximation. Among these, the \emph{functional maps} approach has grown popular due to its high efficiency and the smoothness of the resulting correspondences~\cite{ovsjanikov2012functional}.

Instead of searching directly for a correspondence $\pi$ between two surfaces $\M$ and $\N$, the functional maps approach aims at finding a linear map $T_{\pi} : \FN \to \FM$, induced by $\pi$, from the space $\FN$ of real-valued functions over $\N$ to the space $\FM$ of real-valued functions over $\M$. Given any function $f : V_{\N} \to \Reals$, the map $T_{\pi}$ gives the corresponding function $T_{\pi}(f) : V_{\M} \to \Reals$ that maps $v_i \mapsto f(\pi(v_i))$, for each $v_i \in V_{\M}$.

Given two sets of (orthogonal) basis functions $\evecs{\M}, \evecs{\N}$ for the spaces $\FM,\FN$, respectively, a function $f : V_{\N} \to \Reals$ can be expressed as a linear combination of $\evecs{\N}$
\begin{equation}
    f
    =
    \sum_{i = 0}^{\infty}
    \alpha_i \phi_{\N}^{(i)}
    =
    \sum_{i = 0}^{\infty}
    \langle f, \phi_{\N}^{(i)}\rangle_{\N} \phi_{\N}^{(i)}\,,
\end{equation}
where $\phi_{\N}^{(i)}$ is the $i$-th basis function in $\evecs{\N}$ and $\langle \cdot, \cdot\rangle_{\N}$ denotes the inner product between functions over $\N$. The coefficients $\alpha_i$ denotes the projection of $f$ onto the basis functions $\phi_{\N}^{(i)}$. Usually, $\evecs{\M}, \evecs{\N}$ are chosen to be the eigenfunctions of the Laplace-Beltrami operator on $\M$ and $\N$.
Since $T_{\pi}$ is linear, the mapping of $f$ onto $\M$ is given by
\begin{equation}
    T_{\pi}(f)
    =
    \sum_{i = 0}^{\infty}
    \alpha_i T_{\pi}(\phi_{\N}^{(i)})
    =
    \sum_{i, j = 0}^{\infty}
    \alpha_i c_{i, j} \phi_{\M}^{(j)}\,,
\end{equation}
where $c_{i, j} = \langle T_{\pi}(\phi_{\N}^{(i)}), \phi_{\M}^{(j)} \rangle_{\M}$. By truncating the two bases to $k$ functions, the coefficients $c_{i, j}$ can be stored into a matrix $\mathbf{C} \in \Reals^{k \times k}$, leading to the matrix equation
\begin{equation}
    T_{\pi}(f)
    \approx
    \evecs{\M} \mymat{C} \evecs{\N}^{\dagger} \myvec{f}\,,
\end{equation}
where $^{\dagger}$ denotes the Moore-Penrose pseudo-inverse.
Notably, $T_{\pi}(\evecs{\N}) \approx \evecs{\M}\mymat{C}$, thus by knowing the matrix $\mymat{C}$ it is possible to extract the unknown correspondence $\pi$ via a nearest-neighbor search, $\mathrm{NN}(\evecs{\N}, \evecs{\M}\mymat{C})$.


\section{Method}\label{sec:method}

The functional maps framework is not bound to the domain of 2-dimensional surfaces. Indeed, the Laplace-Beltrami Operator (LBO) can be defined on Riemannian manifolds of arbitrary dimension, and it always admits a spectral decomposition. We discretize the LBO of a tetrahedral mesh $\M = (V_{\M}, T_{\M})$ using the $n$-dimensional cotangent formula~\cite{liao2009gradient,crane2019n} which defines the discrete LBO as a pair of matrices $\mymat{S}$ and $\mymat{W}$, denoted as \emph{stiffness} and \emph{mass} matrices, respectively. Its spectral decomposition can be obtained by solving the generalized eigenproblem $\mymat{S}\evecs{} = \mymat{W}\evecs{}\evals{}$. 

Implementation details and a discussion on the choice of the LBO are provided in Appendix~\ref{sec:laplacian}. We highlight that most applications we address involve analyzing information at the surface (\ie, the boundary of a tetrahedral mesh), compelling us to impose Neumann's boundary conditions to prevent zero-valued eigenfunctions on the surface.

\begin{figure}[!t]
    \centering
    \includegraphics[width=0.9\columnwidth]{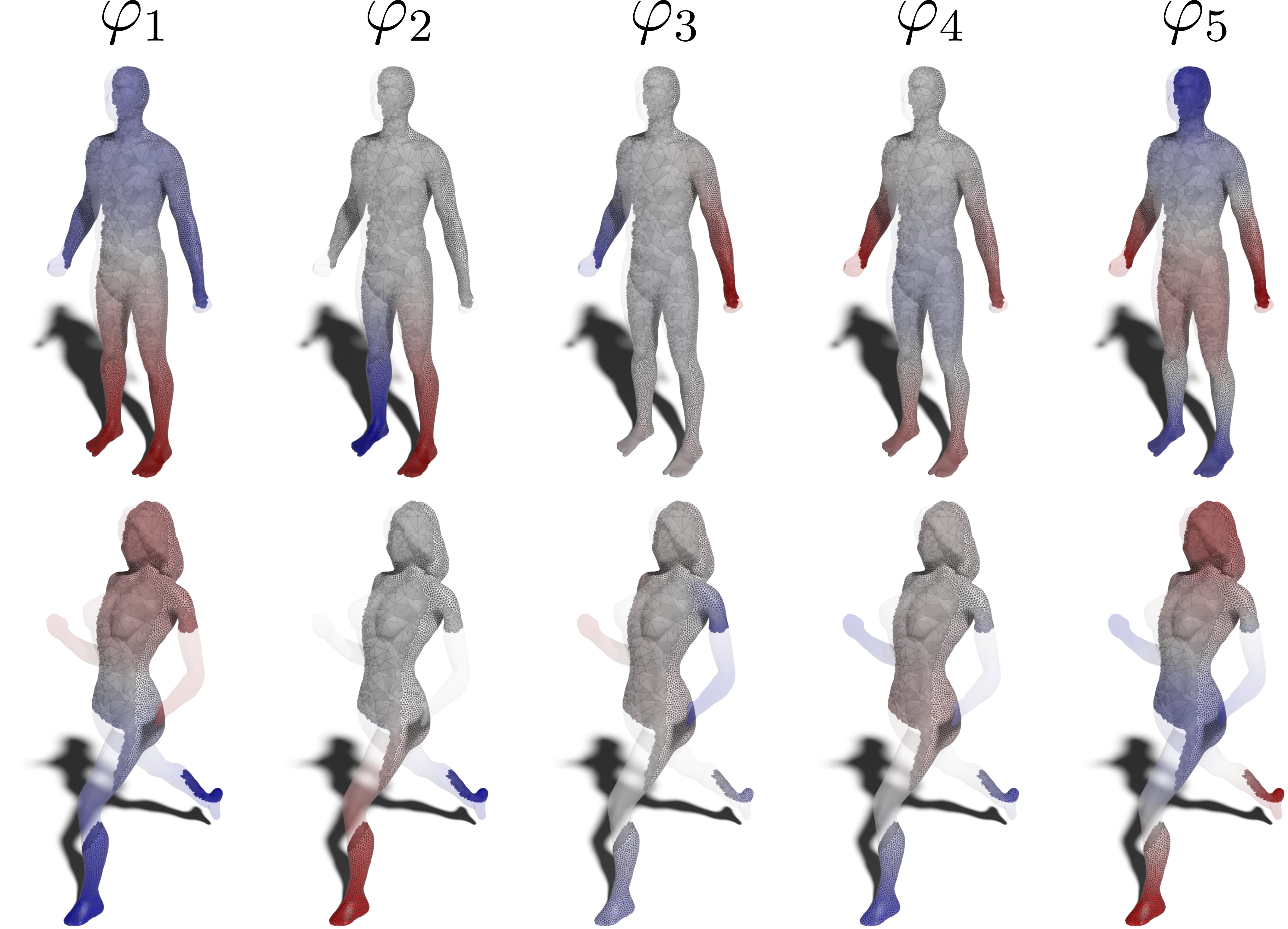}
    \caption{The first 5 non-constant volumetric LBO eigenfunctions on two humanoid non-isometric shapes}
    \label{fig:lbo-humans-compare}
\end{figure}

Most successful algorithms for estimating the functional map between shapes rely on the property that intrinsic structures and processes behave similarly on similar surfaces. For instance, the invariance of the LBO basis and the heat diffusion under (quasi-)isometric deformations plays a key role in stable functional map estimation~\cite{nogneng2017informative,nogneng2018improved,ovsjanikov:2016:fmapsmatching}. Despite this invariance holds for continuous volumetric shapes (see \cref{fig:lbo-humans-compare}), in a discrete setting the position and connectivity of the vertices can heavily impact the numerical stability of the Laplacian operator~\cite{lescoat2020spectral}. 
Appendix~\ref{app:eigencompare} provides a detailed empirical study on the invariance of the LBO in the volumetric setting, and Appendix~\ref{sec:volisometries} contains a deeper discussion on isometric deformations between volumes.

\subsection{Spectral Connectivity Transfer}\label{sec:method:volmap}

The success of functional mapping is also due to its ability to smooth out signals and correspondences: a truncated basis filters out high-frequency details, producing smoothly interpolated signals. This smoothness, combined with the fact that the basis is usually much smaller than the number of vertices, allows to approximate the spectral representation of noisy and partial signals. This feature has been exploited by~\citet{melzi2019zoomout} to efficiently approximate a functional map using only a reduced sampling of the original surface. In our setting, we take advantage of this property to transfer connectivity between volumes.

\paragraph*{Functional Connectivity Transfer.}
Let $\M = (V_{\M}, T_{\M})$ be a tetrahedral mesh with surface $\partial\M = (V_{\partial\M}, F_{\partial\M})$. Let $\partial\N = (V_{\partial\N}, F_{\partial\N})$ be another surface in known correspondence with $\partial\M$ through the (surface) map $\pi : \partial\M \to \partial\N$. There are various options to obtain a tetrahedralization $\N$ of the interior of $\partial\N$~\cite{hang2015tetgen,diazzi2023constrained}. However, by no means, its connectivity will match the connectivity of $\M$.

Given a surface $\partial\N$ with induced tetrahedral mesh $\N$, we exploit the known correspondence $\pi : \partial\M \to \partial\N$ to induce a volumetric correspondence $\pi' : \M \to \N$. Let $\evecs{\M}, \evecs{\N}$ be the LBO eigenbases of the volumes $\M$ and $\N$, respectively, and let us denote with $\evecs{\partial\M} = (\evecs{\M})_{\partial}, \evecs{\partial\N} = (\evecs{\N})_{\partial}$ their restrictions to the surfaces.
We stress that $\evecs{\partial\M}$ is not the LBO eigenbasis of $\partial\M$. The functional map $\mymat{C}$ induced by $\pi'$ must be such that
\begin{equation}\label{eq:fmap-volmap}
\evecs{\M}\mymat{C}
=
T_{\pi'}(\evecs{\N})\,.
\end{equation}

We consider that the rows of $\evecs{\partial\M}$ (resp. $\evecs{\partial\N}$) are a subset of the rows of $\evecs{\M}$ (resp. $\evecs{\N})$. Recalling that $\pi$ is the restriction of $\pi'$, we get
\begin{equation}
\evecs{\partial\M}\mymat{C}
=
(\evecs{\M}\mymat{C})_{\partial}
=
(T_{\pi'}(\evecs{\N}))_{\partial}
=
T_{\pi}(\evecs{\partial\N})\,.
\end{equation}
In general, the size of the basis (\ie, the number of columns of $\evecs{\partial\M}$) is significantly smaller than the number of vertices (\ie, the number of rows), even if we only consider the vertices at the surface. Intuitively, this suggests that the restricted eigenfunctions $\evecs{\partial\M}$ at the surface are linearly independent. Indeed, classical results from control theory support this intuition, proving that the boundary traces (\ie, the restriction to the boundary) of the eigenfunctions of any elliptic operator are linearly independent in a system with Neumann boundary conditions~\cite{lasiecka1983stabilization,triggiani2008linear} (see the Appendix~\ref{sec:boundarytrace} for a deeper analysis of the descriptive power of the boundary traces).
Therefore, $\evecs{\partial\M}$ admits left inverse and the volumetric functional map can be approximated via the known surface correspondence as $\mymat{C} \approx \evecs{\partial\M}\pinv T_{\pi}(\evecs{\partial\N})$.

The functional map $\mymat{C}$ can then be used with the full spectrum to transfer the coordinates of $\N$ onto $\M$. Namely
\begin{equation}\label{eq:coord-tran}
\begin{gathered}
    T_{\pi'}(x_{\N}) = \evecs{\M}\mymat{C}\evecs{\N}\pinv x_{\N}\,,
\end{gathered}
\end{equation}
and similarly for $y_{\N}$ and $z_{\N}$.
By using the values $T_{\pi'}(x_{\N}), T_{\pi'}(y_{\N}), T_{\pi'}(z_{\N})$ as coordinates for the vertices of $\M$, we get a connectivity transfer from $\M$ onto $\N$.

\begin{figure}[t]
    \centering
    \includegraphics[width=\columnwidth]{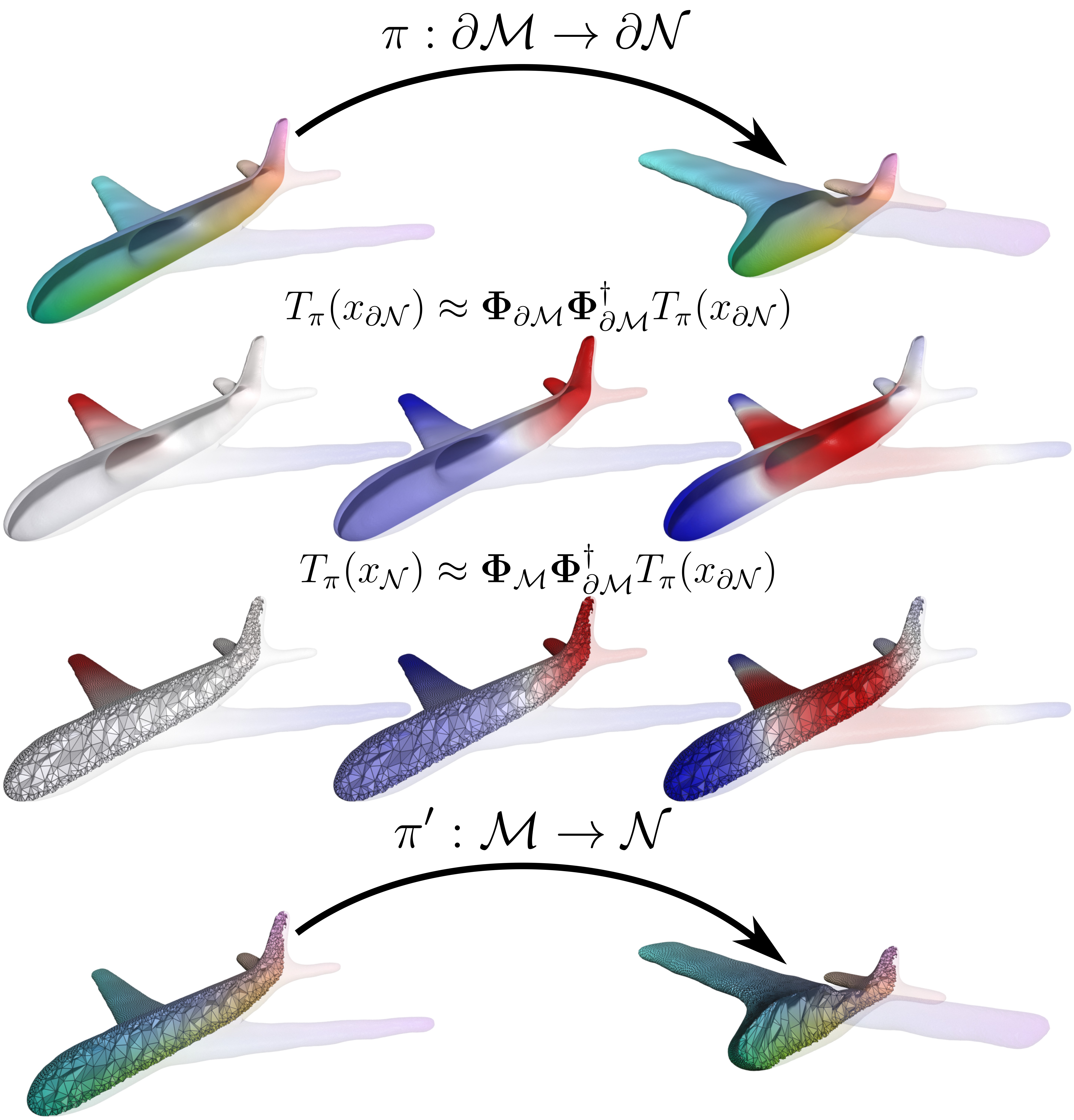}
    \caption{Our pipeline for extrapolating interior coordinates from surface correspondences. Given the surface map $\pi$ (first row), we approximate the spectral embedding of the surface coordinates of $\partial\N$ using the boundary restriction of the eigenfunctions of $\M$ (second row). Using the eigenfunctions on the entire volume, we reconstruct the interior coordinates from the spectral embedding (third row), and use these coordinates to transfer the inner connectivity (fourth row).}
    \label{fig:extrapolation-explain}
\end{figure}

\paragraph*{Spectral Extrapolation of Coordinates.}
A watertight surface already encodes information about the enclosed volume. Since the LBO provides smoothness guarantees in signal reconstruction, it is a good candidate for extrapolating the interior coordinates of a volume from its surface.

Given the ground truth correspondence $\pi : \partial\M \to \partial\N$, we can transfer the surface coordinates $x_{\partial\N}, y_{\partial\N}, z_{\partial\N}$ of $\partial\N$ onto $\partial\M$ as $T_{\pi}(x_{\partial\N}), T_{\pi}(y_{\partial\N}), T_{\pi}(z_{\partial\N})$. We then consider the LBO eigenbasis $\evecs{\M}$ and its restriction $\evecs{\partial\M}$ to the surface. Following a reasoning analogous to the discussion in the previous paragraph, we have that
\begin{equation}
\begin{gathered}
    (T_{\pi'}(x_{\N}))_{\partial}
    \approx
    (\evecs{\M}\evecs{\M}\pinv T_{\pi'}(x_{\N}))_{\partial}
    =
    \evecs{\partial\M}\evecs{\M}\pinv T_{\pi'}(x_{\N})\,,
    \\
    T_{\pi}(x_{\partial\N})
    \approx
    \evecs{\partial\M}\evecs{\partial\M}\pinv T_{\pi}(x_{\partial\N})\,.
\end{gathered}
\end{equation}
Since $(T_{\pi'}(x_{\N}))_{\partial} = T_{\pi}(x_{\partial\N})$, then it must also be $\evecs{\M}\pinv T_{\pi'}(x_{\N}) \approx \evecs{\partial\M}\pinv T_{\pi}(x_{\partial\N})$.
Finally, the coordinates on the interior of $\N$ can be extrapolated by reconstructing with the full basis:
\begin{equation}\label{eq:coord-extra}
\begin{gathered}
    T_{\pi'}(x_{\N}) = \evecs{\M}\evecs{\partial\M}\pinv T_{\pi}(x_{\partial\N})\,,
\end{gathered}
\end{equation}
and analogously for $y_{\N}$ and $z_{\N}$.
A visual representation of our pipeline is shown in \cref{fig:extrapolation-explain}.



Concrete examples of reliable connectivity transfer using both a functional transfer of interior coordinates and their extrapolation from the spectral embedding of the surface coordinates are depicted in~\cref{fig:volmap-cactus} (resp. \emph{Transfer} and \emph{Extrapolation}). For this specific test, we extended the LBO eigenbasis with the Coordinates Manifold Harmonics (CMH) basis~\cite{marin2019cmh}, which also includes three additional functions encoding the extrinsic per vertex $xyz$ coordinates and is known to be better suited to reconstruct extrinsic information. Additional experiments on the whole dataset, considering both the LBO and CMH eigenbasis are reported in \cref{ssec:conn_trans}.


\subsection{Volume-Aware Surface Correspondence}\label{sec:method:surfmap}

Besides purely volumetric applications, we observe that a good functional map among volumes also induces a good functional map among their external surfaces. Intuitively, a volumetric shape is a richer structure than a surface, yielding more information both locally and globally. Therefore, we guess that functional maps among tetrahedral meshes are more accurate and informative than maps among triangular meshes. This idea is grounded from the results of \citet{raviv2010volumetric}, who have shown that using volume-informed shape descriptors has positive effects in computing correspondences between surfaces.
As mentioned in \cref{sec:method:volmap}, the functional map that relates the LBO eigenbases of two volumetric shapes also relates their restriction to the surfaces. Consequently, the volumetric map can be used to infer a correspondence between surfaces.

Given two surface meshes $\partial\M, \partial\N$, we compute the tetrahedral meshes $\M, \N$ representing their interior volume. 
We then compute the functional map $\mymat{C}$ relating their eigenbases via $\evecs{\M}\mymat{C} = T_{\pi}(\evecs{\N})$, and we can use that same functional map for defining the surface relation $(\evecs{\M})_{\partial}\mymat{C} = (T_{\pi}(\evecs{\N}))_{\partial}$. The surface correspondence is eventually extracted via nearest neighbor search, $\mathrm{NN}(\evecs{\partial\M}\mymat{C}, \evecs{\partial\N})$.

\section{Results and Applications}\label{sec:results}

We implemented our tool in \textsc{Matlab}, extending the surface only implementation of the functional map from~\cite{nogneng2017informative} and ZoomOut~\cite{melzi2019zoomout}. Our code is available at 
\url{https://github.com/filthynobleman/vol-fmaps}.

\subsection{Volumetric Functional Maps}
\label{ssec:res_vol_map}

\begin{figure}[t]
    \centering
    \includegraphics[width=\columnwidth]{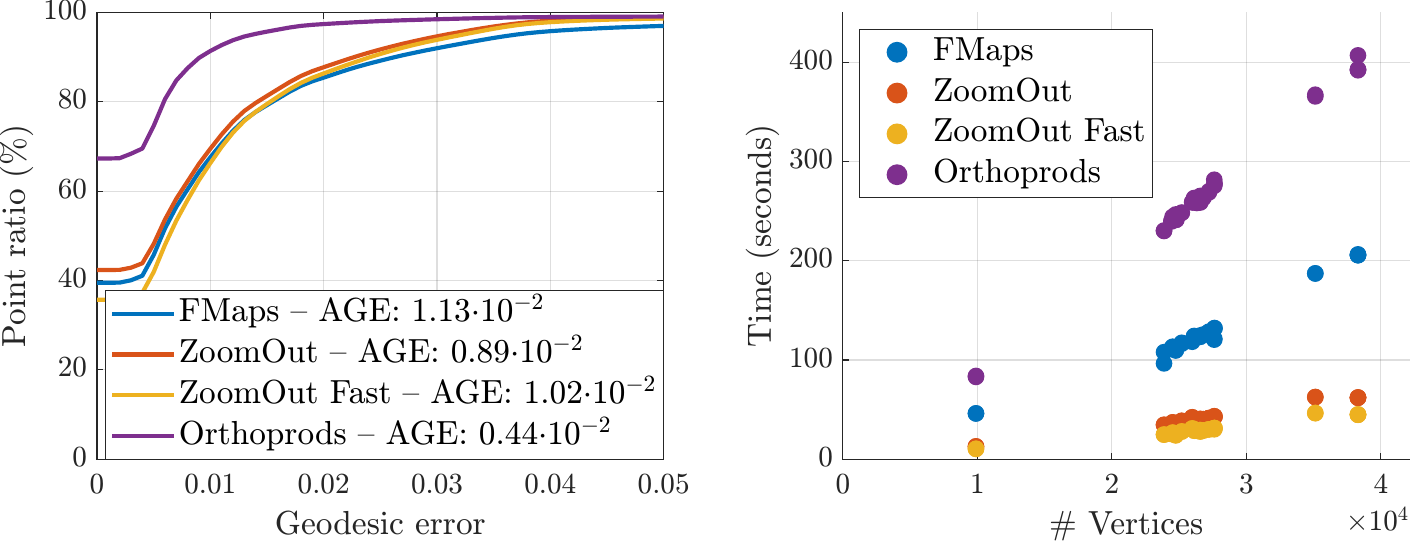}
    \caption{Left: average geodesic error curves on the pairs from Su \etal~\cite{su2019practical} resulting from the application of FMaps~\cite{nogneng2017informative}, ZoomOut~\cite{melzi2019zoomout}, and Orthoprods~\cite{maggioli2021orthogonalized}. Right: runtime for the three methods plotted against the number of vertices in the source mesh.
    }
    \label{fig:fmap-volume-res}
\end{figure}

\begin{figure}[t]
    \centering
    \includegraphics[width=\linewidth]{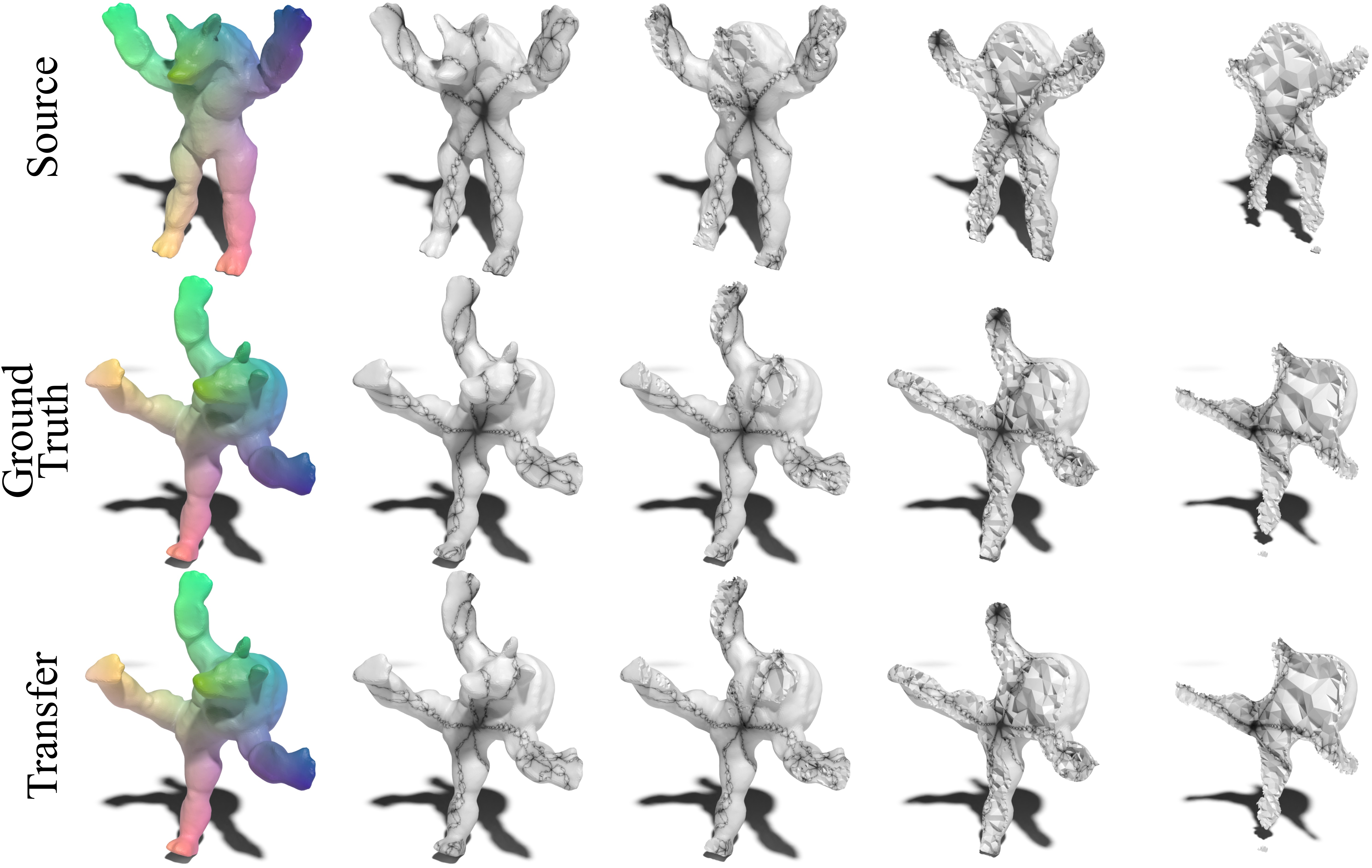}
    \caption{Coordinate transfer between two volumetric shapes using the Orthoprods basis. The normalized coordinates are treated as RGB channels for visualization (first column). For further validation, the transferred coordinates are also used for generating an error-sensitive procedural texture, which is visualized on different slices of the volume (second to fourth columns). 
    }
    \label{fig:newton}
\end{figure}

We validate our prototype and assess its ability to devise high-quality volume maps by comparing our correspondences against the ground truth. Unlike the surface setting, where a constantly growing amount of public data has become available for validation, almost no volumetric datasets are available for this specific task. We consider a set of 40 pairs of volume meshes with identical connectivity, which thus define an injective piece-wise linear map as described in~\cref{ssec:pl_maps}. This dataset, originally proposed by~\cite{su2019practical} and then used to validate~\cite{,du2020lifting}, covers diverse classes of objects, such as: animals, humanoids, hands, and a variety of man-made objects like glasses, lamps, laptops and scissors. For each pair of objects, we compute a volumetric functional map and extract a point-to-point correspondence as described in \cref{sec:method}. Then, for each vertex $v$, we measure the geodesic error as the geodesic distance $d_{\N}(\pi_{\mathrm{gt}}(v), \pi(v))$ between its mapped point $\pi(v)$ and its ground truth counterpart $\pi_{\mathrm{gt}}(v)$. In \cref{fig:fmap-volume-res} we show the cumulative curves of the geodesic error obtained with different functional map estimation algorithms, as well as their runtime in relation to the vertex count. We also provide the average geodesic error (AGE) in the legend of the figure. For comparisons, we use the well-established algorithm based on descriptor preservation (FMap)~\cite{nogneng2017informative} and ZoomOut~\cite{melzi2019zoomout}, using both the standard and the fast implementation. For the fast implementation, we use the surface points as samples to validate the claim in \cref{sec:method:volmap} that the surface encodes information about the enclosed volume. We also test our framework replacing the LBO eigenbasis with the orthogonalized eigenproducts basis (Orthoprods) proposed by~\citet{maggioli2021orthogonalized}, but instead of using the approach described by the authors for the functional map estimation, we plug the basis into the ZoomOut pipeline. As expected, the extended Orthoprods basis has higher descriptive power, but this comes with additional computational cost for aligning the larger bases. In contrast, ZoomOut achieves the best trade-off between accuracy and runtime. Overall, our results demonstrate that the volumetric functional map framework produces high-quality maps, matching the performance obtained in the surface setting. This can be further appreciated in the signal transfer example from \cref{fig:newton}, where we transfer the coordinates function between two meshes with the Orthoprods bases and use the resulting values to compute an error-sensitive procedural texture~\cite{maggioli2022newton}. 
All values reported in this subsection correspond to volumetric extensions of the considered metrics. A more detailed evaluation of the maps is provided in Appendix~\ref{sec:volumemetrics}.

\begin{table}[t]
    \centering
\caption{Average percentage of flips using the methods described in \cref{sec:method:volmap} at varying the basis size. The average runtime (in seconds) to compute the largest basis considered is also reported.}    
    \resizebox{\columnwidth}{!}{%
	{\footnotesize
		\begin{tabular}{cccccc}
			Method  & 5\% eigs & 10\% eigs & 15\% eigs & 20\% eigs & Time (s)
			\\
			\hline
			Transfer (LBO)         & 4.63\% & 1.50\% & 1.18\% & 0.93\% & 1127
			\\
			Transfer (CMH)         & 4.75\% & 0.84\% & 0.45\% & 0.34\% & 1127
			\\
			Extrapolation (LBO)    & 4.39\% & 1.45\% & 1.15\% & 0.92\% & 554
			\\
			Extrapolation (CMH)    & 4.90\% & 0.99\% & 0.56\% & 0.42\% & 554
		\end{tabular}
	}
    }
	\label{tab:connectivity-transfer}
\end{table}

\begin{figure}[t]
	\centering
	\includegraphics[width=\columnwidth]{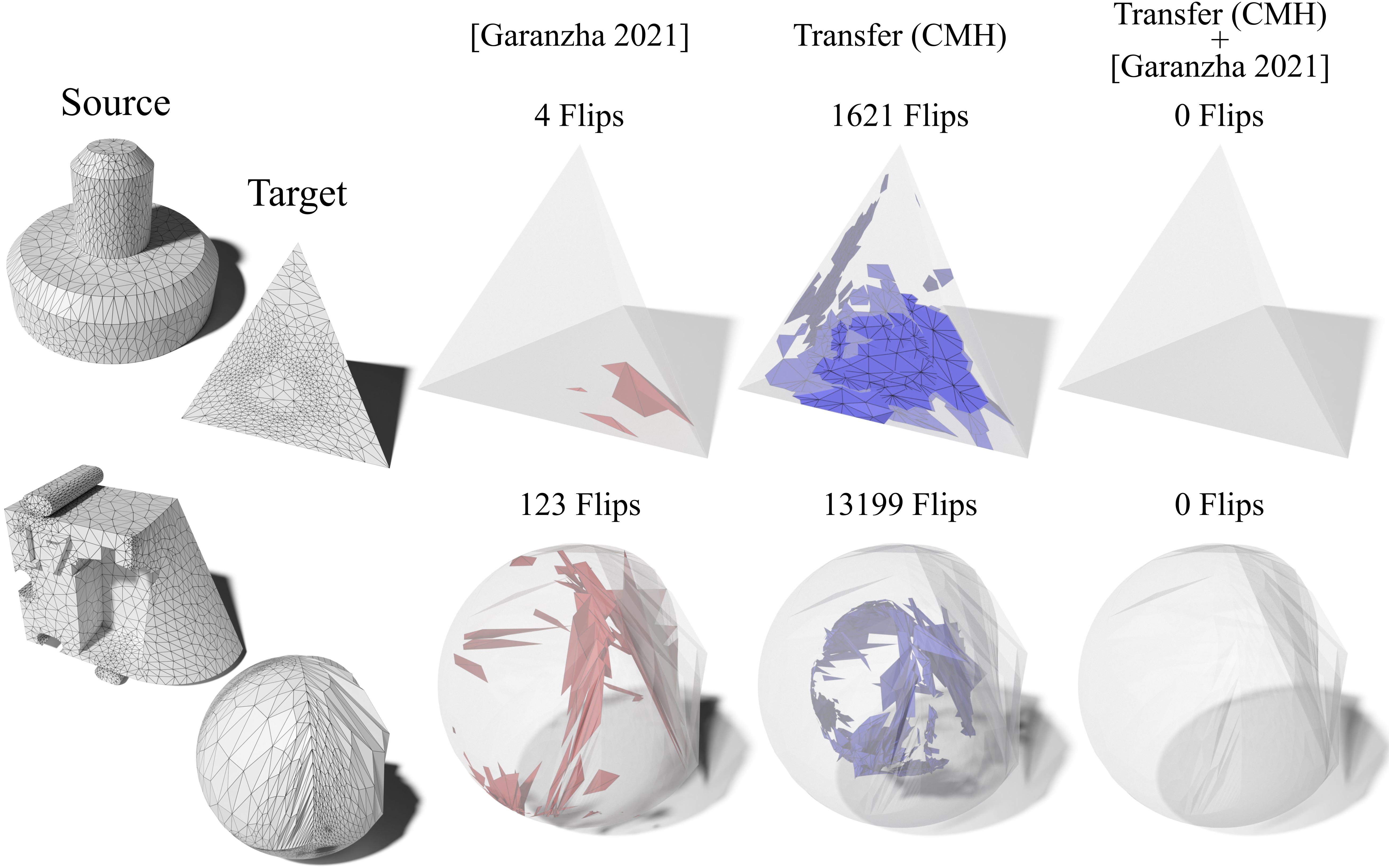}
	\caption{Left: two of the 1405 failure cases of the untangling algorithm proposed by \citet{garanzha2021foldover}, which emerged from the challenging large scale validation described in \cite[\S6.3]{nigolian2023expansion}. Right: computing an initial connectivity transfer with our tool to warm start the same algorithm allows to retain a fully bijective map, thus improving the overall robustness.}
	\label{fig:fix-flip}
\end{figure}

\subsection{Connectivity Transfer}
\label{ssec:conn_trans}
Representing volume maps using two tetrahedral meshes that share the same connectivity but have different vertex coordinates is a common choice for many practical applications involving digital fabrication~\cite{liu2024neural,zhang2022s3,etienne2019curvislicer}, medicine~\cite{abulnaga2021volumetric} and hexahedral meshing~\cite{pietroni2022hex,bruckler2023collapsing}. As detailed in~\cref{sec:method:volmap}, our framework can be used to transfer the connectivity of a given volume mesh onto a target domain, either by using a functional map to transfer the vertex coordinates (\S\ref{sec:method:volmap}~\emph{Functional Connectivity Transfer}) or by extrapolating the connectivity from its outer surface (\S\ref{sec:method:volmap}~\emph{Spectral Extrapolation of Coordinates}). 

\begin{figure}[t]
	\centering
	\includegraphics[width=\linewidth]{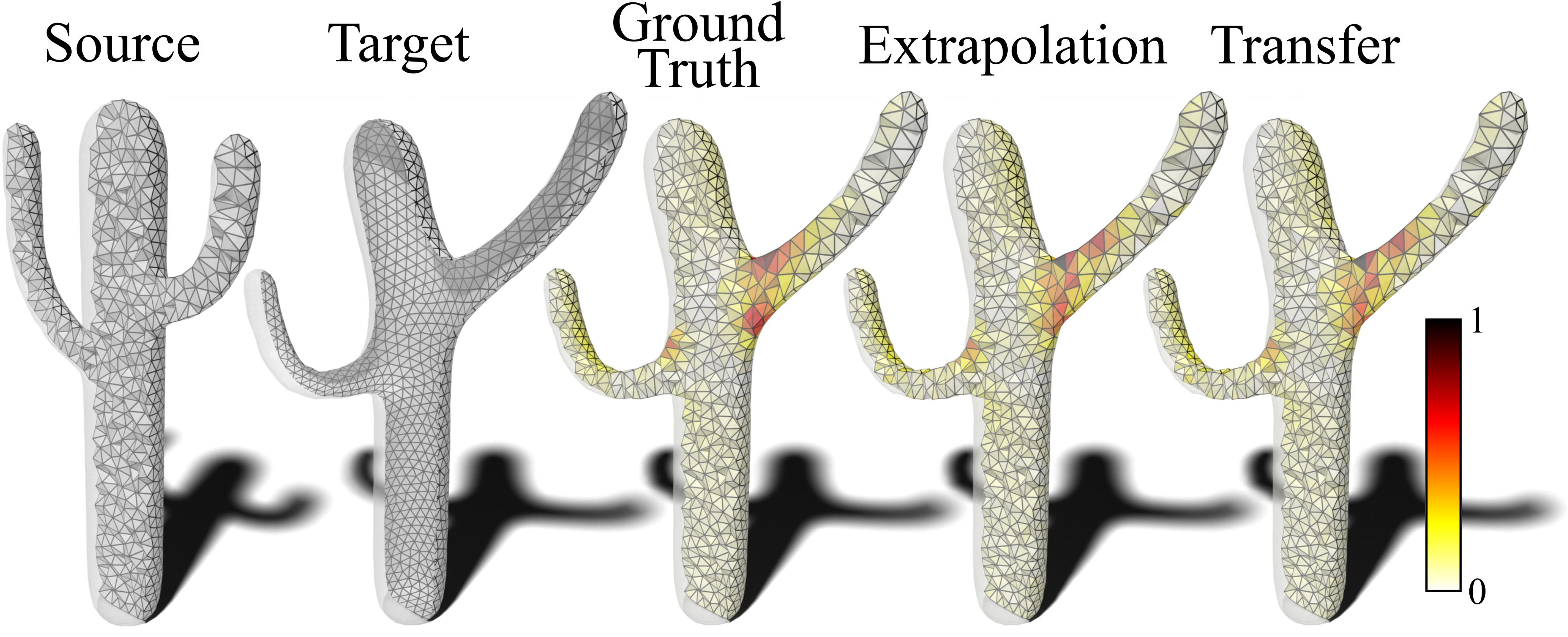}
	\caption{From left to right: tetrahedral mesh; surface of the target domain; ground truth connectivity transfer; connectivity transfer obtained by extrapolating the interior vertices from the surface coordinates; connectivity transfer obtained by transferring the coordinates with the functional map. Each tetrahedron $t$ is colored according to the geometric distortion of the map, measured as $|1 - \det(J_t)|$, where $J_t$ is the Jacobian of the affine transformation mapping $t$ from the source to the target. All maps shown in this figure are fully injective (i.e., $\forall t, \det(J_t)>0$).}
	\label{fig:volmap-cactus}
\end{figure}

If the whole spectrum is considered, the LBO basis spans the entire functional space of the mesh and the functional map is one-to-one~\cite{rodola2017regularized}. Computing the full spectrum is computationally expensive, hence only a fraction of the eigenfunctions are typically computed. We test our prototypes on the dataset of 40 meshes released by \citet{su2019practical} and empirically verify that the amount of inverted elements introduced in the map gracefully decays to zero for growing numbers of eigenfunctions (see Appendix~\ref{sec:volmapbsize} for a detailed analysis). On average, already using $5\%$ of the spectrum the number of flipped cells falls below $5\%$. It consistently decreases as basis grows, and goes below $1\%$ if $20\%$ of the spectrum is considered (\cref{tab:connectivity-transfer}), reaching full injectvity when 25\% of the spectrum is used (\cref{fig:volmap-cactus}). We consider both the LBO spectrum and its extension equipped with three additional functions that encode the extrinsic embedding (CMH). As shown in~\cite{marin2019cmh,MELZI2020-intExt} CMH is better suited to transfer discrete signals such as the surface mesh connectivity. This advantage also manifests on volumes, as reflected in the second and fourth rows of~\cref{tab:connectivity-transfer}. For efficiency, the average running times for computing the larger bases are reported in the last column of the same table. As can be noticed the extrapolation algorithm is roughly $2\times$ faster than volumetric transfer, because in this case only the volumetric spectrum of a single mesh must be computed. 
Note that even using a small fraction of the spectrum, volumetric functional mapping still provides a valuable initial solution for untangling algorithms such as~\cite{garanzha2021foldover,du2020lifting}, which are then required to remove the remaining inverted tetrahedra from the map. This is a standard procedure for generating injective linear maps, where mesh untangling is typically bootstrapped using the Tutte 3D embedding~\cite{hinderink2023galaxy}, which is known to fail on volumes~\cite{alexa2023tutte}. Our tool provides a better initial guess than Tutte. To prove this, we considered the 1405 failures produced in the large-scale validation described in \cite[\S6.3]{nigolian2023expansion}. Substituting the initial guess provided by Tutte with the one produced with our tool, we were able to obtain one-to-one maps in the 85\% of their failure cases, using both LBO and CMH basis functions, dramatically increasing the success rate of this pipeline (\cref{fig:fix-flip}).

\subsection{Segmentation Transfer}
\label{ssec:res_seg_trans}
Statistical shape analysis for medical data primarily benefits from accurate solutions to analyze large collections of shapes~\cite{heimann2009statistical,stegmann2002brief}. The literature shows that the functional maps framework is a prominent tool for medical applications~\cite{melzi2016functional,magnet2023assessing,maccarone2024s4a}. A standard approach to analyze collections of shapes is to determine or build a template surface which acts as a median shape to compare the entire dataset~\cite{huang2019limit}. However, existing approach are limited to the analysis of surfaces, while often medical data are better represented as 3D volumes (\eg, MRI scans~\cite{rinck2019magnetic}). 

\begin{figure}[t]
	\centering
        \vspace{0.5em}
	\includegraphics[width=\columnwidth]{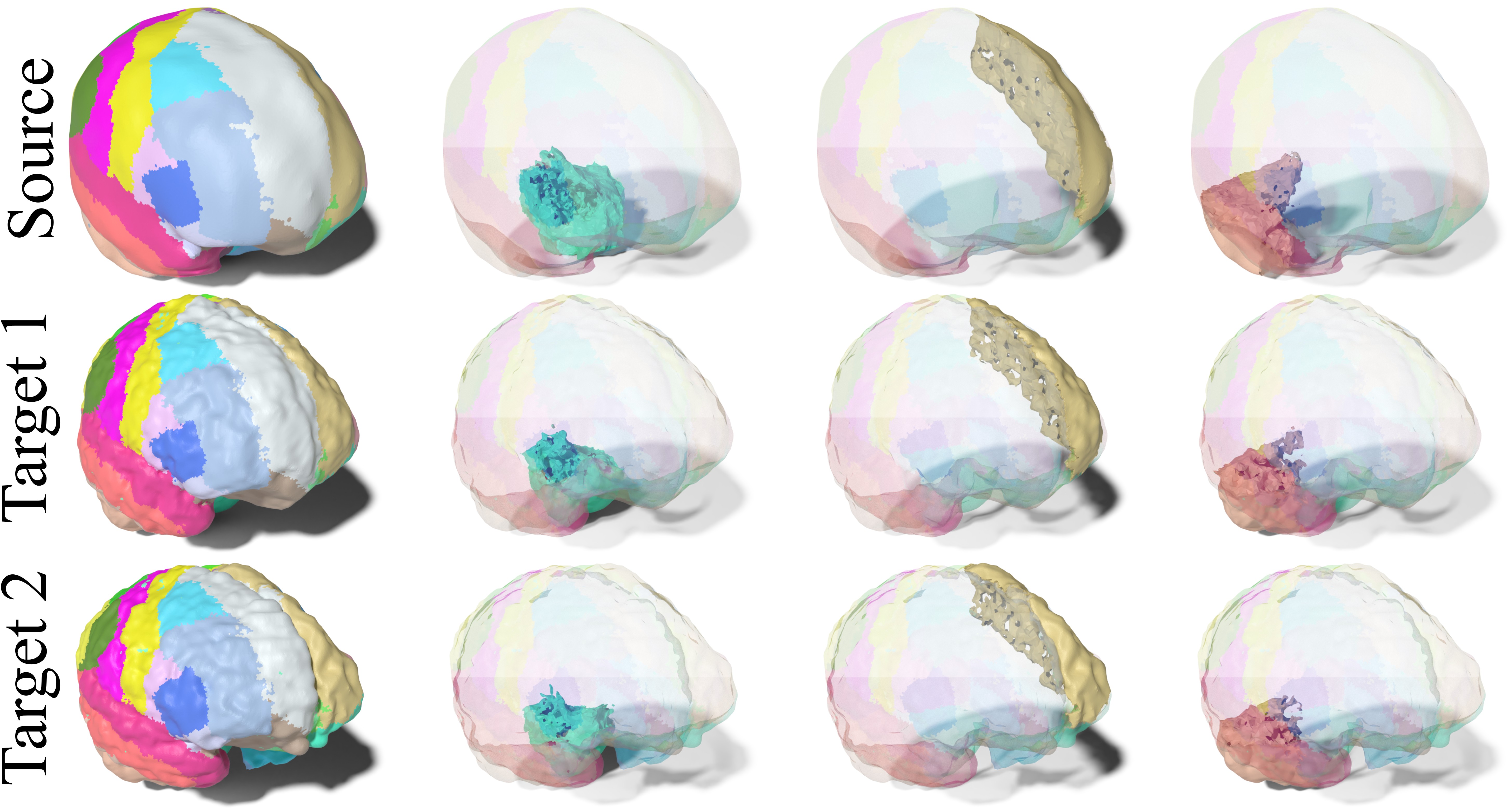}
	\caption{A segmentation of a template brain transferred to two other brains using a correspondence computed with our volumetric functional maps framework. The shapes and the segmentation are from MedShapeNet~\cite{li2024medshapenet}.}
	\label{fig:brains}
\end{figure}

Our approach allows to extend existing methodologies to deal with volumetric data. To prove this, we select two brain shapes from the MedShapeNet dataset~\cite{li2024medshapenet} and a population-averaged template~\cite{evans19933d}. Using TetGen~\cite{hang2015tetgen}, we compute the tetrahedral meshes representing the volumes of the brains, and segment the template using a reference segmentation~\cite{fischl2002whole}. We then compute a volumetric functional map from the template to each shape, extract the point-to-point correspondence and transfer the segmentation. From the results in~\cref{fig:brains}, we can see that, despite the noisy segmentation in the central section, our tool can faithfully transfer the segmentation preserving its overall structure. 

\subsection{Shape Matching}
\label{ssec:res_matching}
Here we study to what extent incorporating volumetric information into the shape matching problem improves accuracy over existing surface-based functional map pipelines.

\paragraph*{Data.} We consider both the outer surfaces of our reference ground truth volumetric dataset (denoted as VOL)~\cite{su2019practical} and well established shape matching datasets, such as SHREC'19~\cite{melzi2019shrec}, SHREC'20~\cite{dyke2020shrec}, and TOPKIDS~\cite{lahner2016shrec}, which we all tetrahedralize using fTetWild~\cite{hu2020fast}. These datasets stress shape matching algorithms 
in different ways, exhibiting geometric and topological noise (SHREC'19), strongly non-isometric matches (SHREC'20) and topological changes (SHREC'20 and TOPKIDS). For SHREC'20, we discard two meshes because the inside/outside system of fTetWild does not reproduce a meaningful volumetric mesh for the comparison. We refer to Appendix~\ref{sec:brokendata} for more details on the data.



\begin{table*}[t]
	\centering
	\resizebox{\linewidth}{!}{%
		{\footnotesize
			\begin{tabular}{c|ccc|ccc|ccc|cc}
    &
    \multicolumn{3}{c|}{LBO} &
    \multicolumn{3}{c|}{CMH} &
    \multicolumn{3}{c}{Orthoprods} \\
    &
    \multicolumn{3}{c|}{\downbracefill} &
    \multicolumn{3}{c|}{\downbracefill} &
    \multicolumn{3}{c}{\downbracefill} &
    \\
    Dataset &
    AGE Surf. &
    AGE Vol. &
    Succ. Rate &
    AGE Surf. &
    AGE Vol. &
    Succ. Rate &
    AGE Surf. &
    AGE Vol. &
    Succ. Rate &
    Slowdown &
    Vertex Ratio\\
    \hline
    VOL & 
    \underline{1.88$\cdot10^{-6}$} &
    2.92$\cdot10^{-6}$ &
    22.50\% &
    \underline{1.78$\cdot10^{-6}$} &
    3.15$\cdot10^{-6}$ &
    20.00\% &
    5.82$\cdot10^{-7}$ &
    \underline{3.19$\cdot10^{-7}$} &
    60.00\% &
    1.92$\times$ &
    1.40$\times$ \\
    
    SHREC'19 &
    1.03$\cdot10^{-1}$ &
    \underline{6.64$\cdot10^{-2}$} &
    53.15\% &
    1.04$\cdot10^{-1}$ &
    \underline{6.56$\cdot10^{-2}$} &
    54.52\% &
    1.11$\cdot10^{-1}$ &
    \underline{7.13$\cdot10^{-2}$} &
    56.42\% &
    2.03$\times$ &
    1.59$\times$ \\

    TOPKIDS &
    1.24$\cdot10^{-1}$ &
    \underline{8.85$\cdot10^{-2}$} &
    60.00\% &
    1.23$\cdot10^{-1}$ &
    \underline{8.86$\cdot10^{-2}$} &
    68.00\% &
    1.17$\cdot10^{-1}$ &
    \underline{8.13$\cdot10^{-2}$} &
    64.00\% &
    2.51$\times$ &
    2.16$\times$ \\

    SHREC'20 &
    \underline{1.82$\cdot10^{-1}$} &
    2.72$\cdot10^{-1}$ &
    14.84\% &
    2.70$\cdot10^{-1}$ &
    \underline{1.82$\cdot10^{-1}$} &
    55.49\% & 
    \underline{1.83$\cdot10^{-1}$} &
    3.01$\cdot10^{-1}$ &
    8.24\% &
    1.01$\times$ &
    0.80$\times$
\end{tabular}
		}
	}
	\caption{Results of each functional maps implementation on each dataset. We report geodesic error averaged across the dataset (AGE) and percentage of single pairs where our approach obtains a better AGE (Succ. Rate). Computational overhead strongly correlates with mesh growth (right). We report: average slowdown induced by volumes and average vertex ratio between surfaces and volumes.}
	\label{tab:dataset-methods}
\end{table*}

\paragraph*{Comparative Analysis.} 
We validate the approach in \cref{sec:method:surfmap} by comparing it with the standard surface-based pipeline on each dataset. To prove that our volumetric framework can be seamlessly integrated with existing extensions of functional maps, we also test it with different sets of bases, such as Coordinates Manifold Harmonics (CMH)~\cite{marin2019cmh} and Orthogonalized eigenproducts (Orthoprods)~\cite{maggioli2021orthogonalized}. In all cases, we align the bases using the method based on descriptors preservation~\cite{nogneng2017informative} and ZoomOut~\cite{melzi2019zoomout}. For a fair comparison, we use the same bases for both the surface and the volumetric approaches.

The results of our experiments are summarized in~\cref{tab:dataset-methods}, where we show the dataset averaged geodesic error (AGE) for both frameworks and the percentage of single pairs for which our approach obtains a better AGE (Success Rate). We also report the slowdown factor induced by volumes and the ratio between the vertex count of the surface and volume meshes. Interestingly, our approach is consistently more accurate than its surface counterpart, suggesting that incorporating volumetric information into the matching process is always a rewarding choice when a volumetric discretization is available. The increased accuracy comes at the cost of additional computational, due to the introduction of the internal vertices. Inspecting the Slowdown columns in \cref{tab:dataset-methods} we can see that there is a close relationship between the slowdown and the number of vertices introduced, suggesting that the volumetric pipeline is not significantly slower per se, but suffers from the overhead induced by the additional vertices needed to discretize the volume. 

From the scale of the AGE, we can deduce that VOL is not a really challenging dataset for shape matching (watertight shapes, shared connectivity); thus, introducing volumetric information provides a real advantage only with more informative bases, such as Orthoprods. The accuracy of our approach seems related to the choice of the basis also when dealing with significantly more challenging datasets like SHREC'20. This is due to topological errors introduced by tetrahedralization. The topological changes reflect on the LBO eigenbasis and the Orthoprods basis, decreasing the accuracy. The CMH basis uses the vertex coordinates as a regularizer, compensating for the error. We provide a more detailed discussion in Appendix~\ref{sec:matchingbasis}.

\section{Conclusions}\label{sec:conclusions}

We presented the first volumetric functional maps pipeline, showcasing real applications that exploit discrete and continuous signal transfer for solid texturing, connectivity transfer, and segmentation transfer for medical applications, also improving the accuracy of surface matching tasks. Our practical contributions just scratch the surface of this problem and we foresee interesting avenues for future research.

\paragraph*{Limitations.}
Our connectivity transfer shows great potential but is far from being perfect. Indeed, always achieving a fully injective map is technically possible, but considering a large enough portion of the spectrum may easily become computationally prohibitive. Devising smarter techniques to retain more precision for fixed spectrum size a research direction worth investigating.

The higher precision in shape matching tasks comes at the cost of an additional computational effort, because the most expensive steps in the pipeline (eigendecomposition and basis alignment) suffer from the increased vertex count. This limitation is intrinsic to our approach, but it could be mitigated by integrating scalable algorithms for computing and aligning bases specifically designed for volumes.

\paragraph*{Future works.}
A theoretical study of what properties are invariant under a volume functional map must be carried out. Despite our empirical studies show that volumetric maps are more accurate than their surface counterparts, a more profound theoretical analysis would provide useful insight for further developments.


Finally, we emphasize the importance of releasing new volumetric datasets to support ongoing research on volume mapping, and in particular on spectral volume methods. As discussed in \cref{ssec:res_matching}, surface datasets that are well established in the functional map community hardly extend to volumes because of the difficulty to produce tetrahedral meshes of acceptable quality, due to topological and geometric noise, (e.g., large missing parts, overlaps, multiple connected components, non-manifold configurations and self-intersections). While robust volume meshing methods such as~\cite{hu2020fast,diazzi2021convex} can handle intersections with ease, determining semantically plausible separations between the inside and outside portions of the volume in the presence of large gaps remains an open challenge. Recently, a large scale dataset to support ongoing research on piece-wise volume mappings to convex or star-shaped domains has been released~\cite{CL23}. Having similar contributions to support spectral methods would greatly help practitioners and ultimately foster more research in this emerging field.

{
    \small
    \bibliographystyle{ieeenat_fullname}
    \bibliography{biblio}
}

\cleardoublepage

\appendix

\section{Choice of the Laplacian}\label{sec:laplacian}

The original functional maps paper introduced the use of the LBO eigenbasis because it satisfies the desired properties of compactness of representation and stability under near-isometric deformations~\cite[\S5.1]{ovsjanikov2012functional}. Furthermore, later works from \citet{aflalo2015optimality,aflalo2016best} have shown that, in a general setting, and especially when working with smooth signals, the LBO eigenbasis of a manifold $\M$ (of arbitrary dimension) is the optimal orthonormal basis for representing a finite-dimensional subspace of $L^2(\M)$. In this setting, it is reasonable to extend the choice of the LBO to the volumetric domain, especially considering that volumetric deformations do not induce a larger geodesic distortion with respect to near-isometric surface deformations (see \cref{sec:volisometries} for an empirical study in this regard).

On the other hand, it should be noted that discretizing the LBO operator in a way that fulfills all the algebraic properties of the same operator in the smooth setting is often difficult~\cite{wardetzky2007discrete}. In literature there exist multiple alternative discretizations~\cite{jacobson2024optimized,alexa2020properties,mullen2011hot}, each one with its own pros and cons. Among the various alternatives, we chose the cotangent formula due to its popularity, ease of implementation, and efficiency. This choice is not restrictive. Any alternative discretization can be used equivalently. Nonetheless, since most applications that we address involve analyzing information at the surface (which is the boundary of a tetrahedral mesh), we are forced to a discretization that imposes Neumann's boundary conditions to avoid zero-valued eigenfunctions on the surface.

\paragraph*{Volumetric Cotangent Formula}
We discretize the LBO of a tetrahedral mesh $\M = (V_{\M}, T_{\M})$ with the following per vertex relation
\begin{equation}
    \Delta (v_i) =  \sum_{j \in N(v_i)} \!\!w_{ij} (v_i - v_j)\,,
\end{equation}
using the $n$-dimensional cotangent formula~\cite{liao2009gradient,crane2019n} to compute per edge weights $w_{ij}$. These weights consider all tetrahedra $ijkl$ incident to edge $ij$, according to the following formula
\begin{equation}
   w_{ij} = \frac{1}{6} \sum_{ijkl} \vert  v_k - v_l \vert  \cot\theta_{kl}\,, 
\end{equation}
where angle $\theta_{kl}$ denotes the dihedral angle at the edge $kl$ opposite to edge $ij$ w.r.t. tetrahedron $ijkl$.

\section{Spectral Comparison}\label{app:eigencompare}

\paragraph*{LBO Eigenbasis.}
On continuous Riemannian manifolds, the LBO and its eigenbasis are invariant to isometric deformations. However, in the discrete setting, the position and connectivity of the vertices have a significant impact on the equivalence of the operator across different meshes, and changing the triangulation of a surface or its vertex density can strongly affect the spectral decomposition~\cite{lescoat2020spectral}.

Moving from discrete surfaces to discrete volumes makes this problem even more evident. In a regular triangular mesh, each vertex has on average 6 neighboring vertices and 6 neighboring triangles, while in regular tetrahedral meshes, the average size of a vertex neighborhood is 12 vertices and 20 tetrahedra.

\begin{figure}[!t]
    \centering
    \includegraphics[width=0.48\columnwidth]{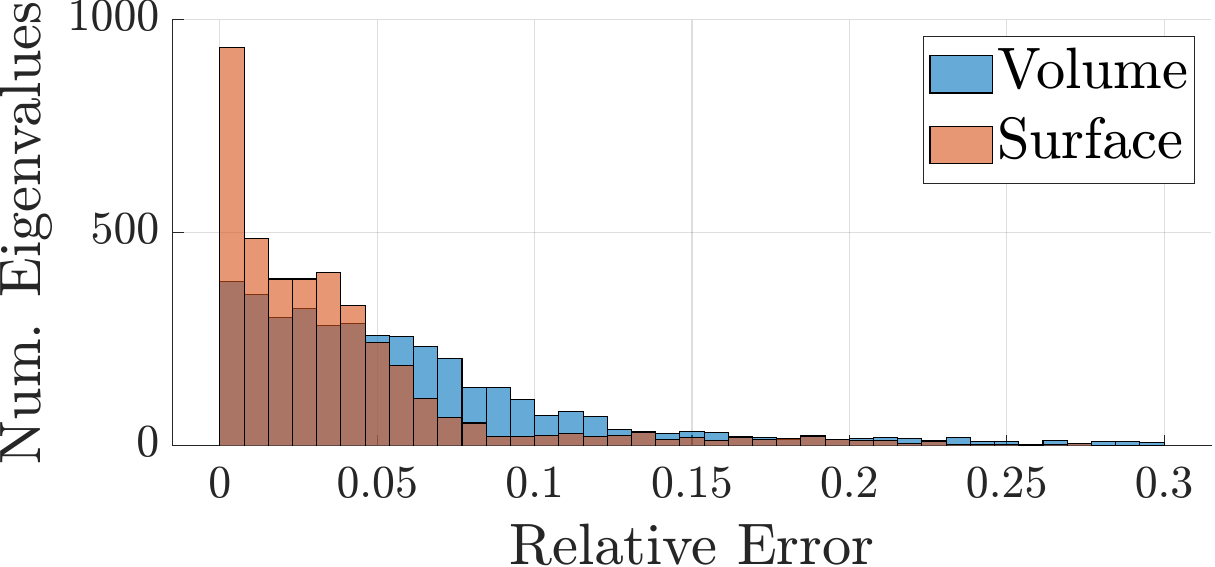}
    \includegraphics[width=0.48\columnwidth]{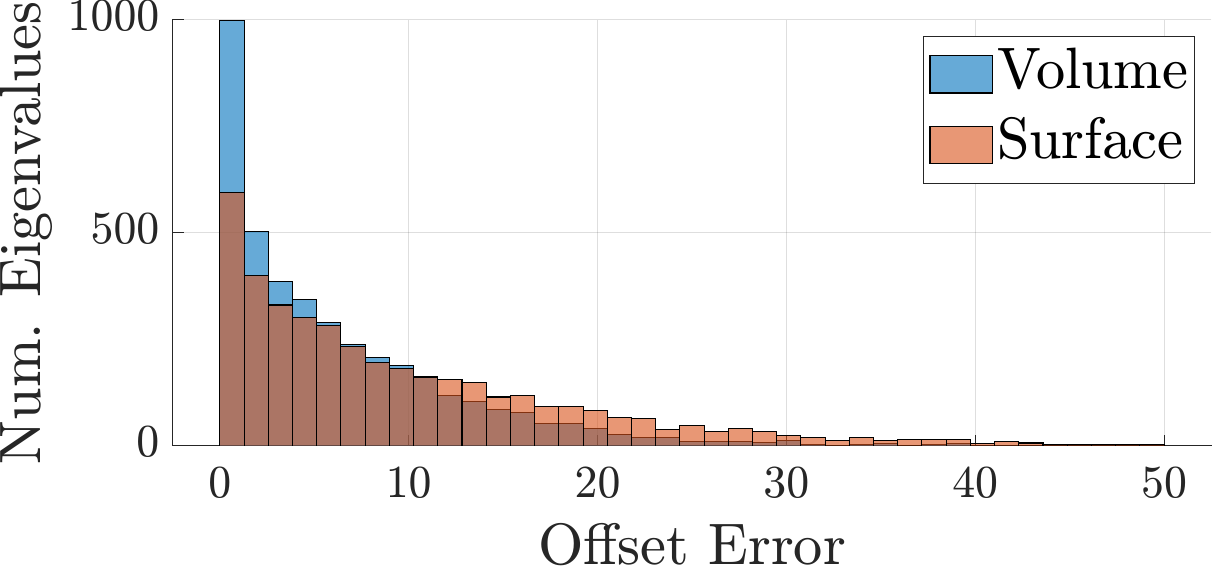}
    \caption{Eigenvalue errors distribution for the volumetric LBO (blue) and the surface LBO (orange).}
    \label{fig:lbo-compare}
\end{figure}

To ensure that the LBO has a consistent behavior across isometric or similar volumetric shapes, we compute the spectral decomposition of the LBO on 40 pairs of shapes with identical connectivity released by \citet{su2019practical} and compare both the eigenvalues and the eigenfunctions of the LBO. Our results are summarized in \cref{fig:lbo-compare}. The first two rows of the figure show that two humanoid shapes share the same LBO eigenfunctions (up to sign flips), even if they are not isometric and do not represent the same subject; the overall similarity is sufficient to generate the same low-frequency spectrum. To compare the eigenvalues, we first compute the vectors $\myvec{\lambda}_{\M}, \myvec{\lambda}_{\N}$ of the first 100 LBO eigenvalues on two tetrahedral meshes. Since the absolute difference between LBO eigenvalues is notoriously not informative, we rely on both the relative difference and the offset difference proposed by \citet{moschella2022learning} (see below).

For each pair of tetrahedral meshes, we also extract the triangular meshes representing their boundary surface and compute the same quantities. For a more meaningful comparison, all the tetrahedral meshes are rescaled to have unitary volume, and all the surface meshes are rescaled to have unitary surface area. The error distributions are shown at the bottom of \cref{fig:lbo-compare}, where we can see that the spectral similarity between pairs of tetrahedral and surface meshes is comparable.

\paragraph*{Offset Error.}
The eigenvalues of the Laplace-Beltrami operator form a non-decreasing sequence that grows with a rate $\lambda_k \in \mathcal{O}(\sqrt[d]{k^2})$, where $d$ is the dimensionality of the manifold~\cite{weyl1911asymptotische}. A direct comparison of the eigenvalues is thus misleading, as the difference will be dominated by the eigenvalues with larger index. One possibility for handling this issue is to introduce relative differences in the comparison. Alternatively, \citet{moschella2022learning} introduced the offset representation for comparing and handling Laplacian spectra. Given two manifolds $\M$ and $\N$, the offset difference can be defined as follows
\begin{gather}
    \mathrm{offset}(\myvec{\lambda}_{\M}, \myvec{\lambda}_{\N}) = | \mathrm{off}(\myvec{\lambda}_{\M}) - \mathrm{off}(\myvec{\lambda}_{\N}) |\,,
    \\
    \mathrm{off}(\myvec{\lambda}_{\M}) = (\mathrm{off}(\lambda_{\M, 1}),\, \mathrm{off}(\lambda_{\M, 2}),\, \cdots\, )\,,
    \\
    \mathrm{off}(\lambda_{\M,i}) = \lambda_{\M,i} - \lambda_{\M,i-1}\,.
\end{gather}

This method for comparing eigenvalues is designed to compensate for the Weyl's law on 2-dimensional manifolds: since $\lambda_k \in \mathcal{O}(k)$, comparing consecutive differences (which asymptotically behave like a constant function) is more informative than comparing absolute values. However, when dealing with 3-dimensional manifolds, the Weyl's law changes to $\lambda_k \in \mathcal{O}(\sqrt[3]{k^2})$ and such consecutive differences are not asymptotically constant anymore. This change to the asymptotic behavior also affects the relative difference: if we consider that the $k$-th eigenvalue can be written as the Weyl's asymptotic relation plus a correction term, namely $\lambda_k = c \sqrt[d]{k^2} + r$, we have that
\begin{equation}
    \frac{c_1\,k + r_1 - c_2\,k - r_2}{c_1\,k + r_1} \not\approx \frac{c_1 \sqrt[3]{k^2} + r_1 - c_2\sqrt[3]{k^2} - r_2}{c_1\sqrt[3]{k^2} + r_1}\,.
\end{equation}
As a consequence neither the relative difference or the offset difference have the same scale for 2-dimensional and 3-dimensional manifolds. 

However, we can use the Weyl's law to compensate the non-linear growth of the eigenvalues. Since $\lambda_k \in \mathcal{O}\sqrt[3]{k^2}$, before computing the offsets we can transform the eigenvalues as $\lambda_k' = \sqrt{\lambda_k^3} \in \mathcal{O}(k)$. With this transformation, the error measures has the same scale for both the surfaces and the volumes. Clearly, this holds under the assumption that all the 2-dimensional manifolds have unitary area and all the 3-dimensional manifolds have unitary volume, which is easy to enforce as a preprocessing step.


\begin{figure}[t]
    \centering
    \includegraphics[width=0.48\columnwidth]{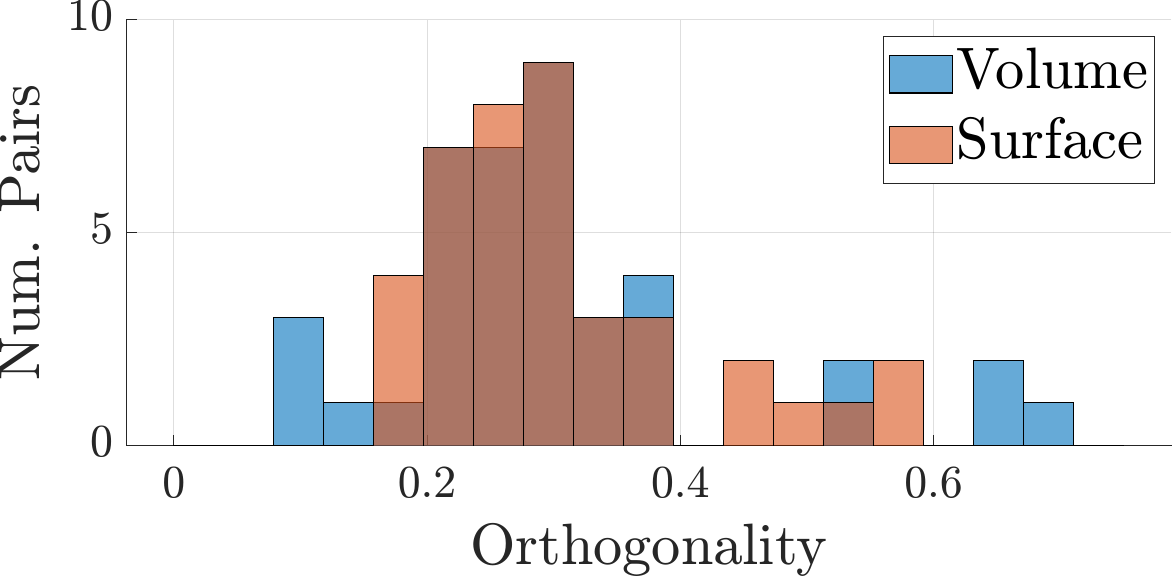}
    \includegraphics[width=0.48\columnwidth]{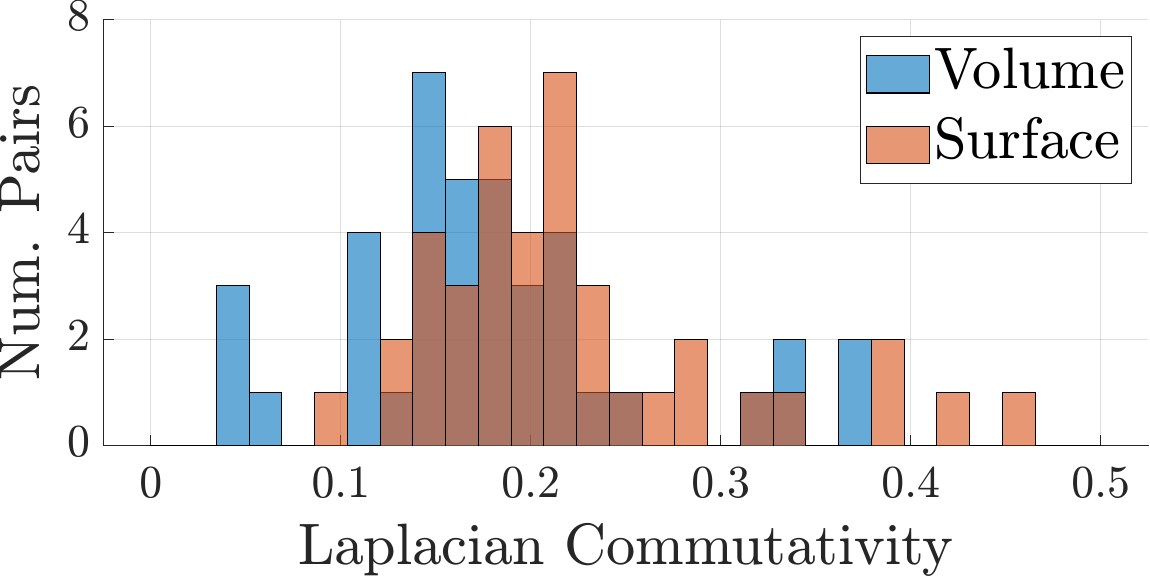}
    \caption{Distribution of the orthogonality (left) and Laplacian commutativity (right) metrics for pairs of volumetric shapes (blue) and surface meshes (orange).}
    \label{fig:fmap-metrics}
\end{figure}

\paragraph{Volumetric Functional Maps.}
Given a definition for the discrete LBO, the definition of the functional maps framework in tetrahedral meshes follows directly from its 2-dimensional counterpart (see \cref{sec:background}). Nonetheless, in order to ensure a seamless integration of the existing approaches in the volumetric setting, we study how the usual properties of a functional map behave on tetrahedral meshes.

Given two meshes $\M$ and $\N$ and a functional map $\mymat{C} : \FM \to \FN$, \citet{lescoat2020spectral} propose two metrics for evaluating the ideal qualities of $\mymat{C}$: the orthogonality $\|\mymat{C}\|_O$ and the Laplacian commutativity $\|\mymat{C}\|_L$
\begin{equation}
    \|\mymat{C}\|_O
    =
    \frac{\|\mymat{C}^\top\mymat{C} - \mymat{I}\|}{\|\mymat{I}\|}\,,
    \qquad
    \|\mymat{C}\|_L
    =
    \frac{\|\mymat{C}\Lambda_{\M} - \Lambda_{\N}\mymat{C}\|}{\|\mymat{C}\Lambda_{\M}\|}\,,
\end{equation}
where $\mymat{I}$ is the identity matrix, $\|\cdot\|$ is the Frobenius norm, and $\Lambda_{\M}, \Lambda_{\N}$ are the diagonal matrices containing the LBO eigenvalues of, respectively, $\M$ and $\N$.

Again, we compute the same quantities for tetrahedral and triangular meshes across the same 40 pairs of shapes, summarizing the results in \cref{fig:fmap-metrics}. As shown in the figure, the quality of volumetric functional maps is comparable to the surface cases.

\section{Volumetric Isometries}\label{sec:volisometries}

\begin{figure*}[t]
    \centering
    \includegraphics[width=\textwidth]{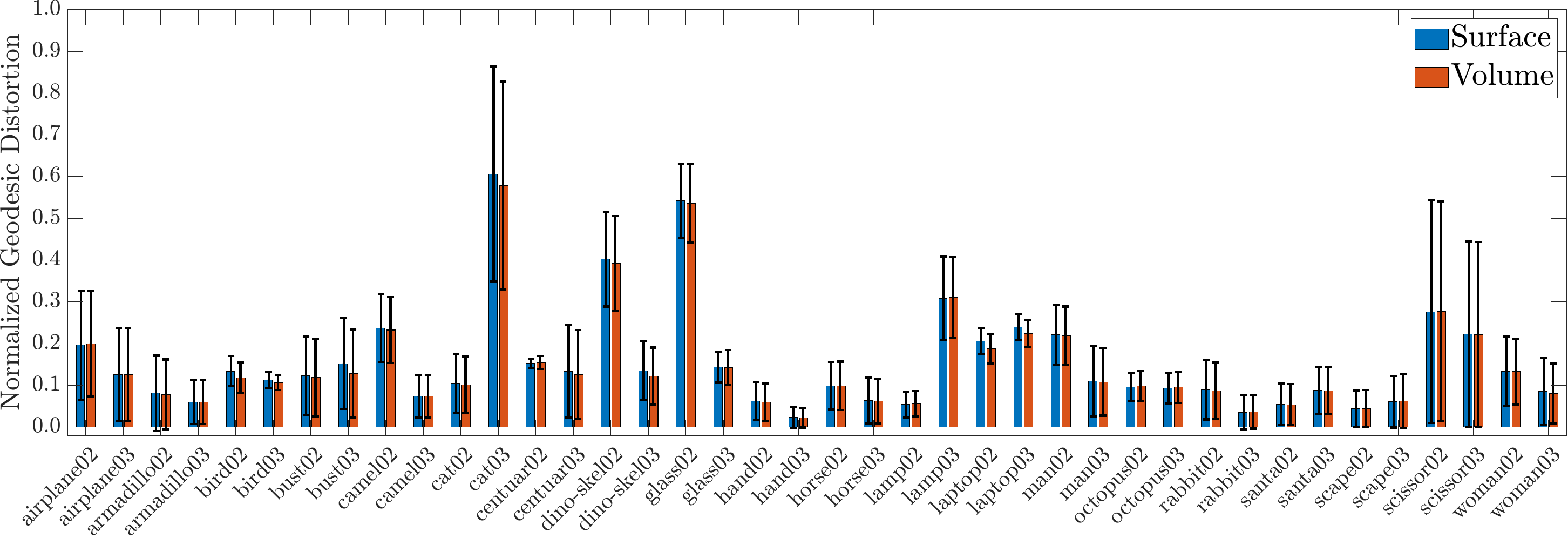}
    \caption{For each subject, the relative geodesic distortion between pairs of volumes (orange) and pairs of surfaces (blue). The plot provides both the average distortion and the standard deviation.}
    \label{fig:geodesic-distortion}
\end{figure*}

The functional maps framework relies on the assumption that the two shapes are isometric or quasi-isometric. In practice, the framework proposed by \citet{ovsjanikov2012functional} works well even under mild non-isometric deformations and meshes at a different scale. However, when dealing with strong non-isometric distortion, more sophisticated approaches are needed, which cannot be easily lifted to manifolds of higher dimensions~\cite{panine22landmark}.

In order to verify that the functional maps framework can be meaningfully transported to a volumetric setting, we need to verify that the level of isometric distortion does not change when a 3D shape is represented as a 2-dimensional surface or a 3-dimensional volume. The dataset introduced by \citet{su2019practical} contains 40 pairs of tetrahedral meshes in bijective correspondence, and thus it is easy to compute the distortion of geodesic distances induced by the correspondence. For each pair of shapes $\M$ and $\N$, we consider 100k random pairs of points $x,y\in\partial\M$. We then compute the geodesic distortion induced by the ground truth correspondence $\pi : \M \to \N$ for both the surface and the volumetric setting, respectively as
\begin{gather}
    \mathrm{SurfErr}(x, y) = \frac{|d_{\partial\M}(x, y) - d_{\partial\N}(\pi(x), \pi(y)|}{d_{\partial\M}(x, y)}\,,
    \\
    \mathrm{VolErr}(x, y) = \frac{|d_{\M}(x, y) - d_{\N}(\pi(x), \pi(y)|}{d_{\M}(x, y)}\,,
\end{gather}
where $d_{\mathcal{S}}$ denotes the approximation of the geodesic distance in the volume $\mathcal{S}$ provided by the Dijkstra distance, and $d_{\partial\mathcal{S}}$ denotes the analogous approximation over the boundary surface $\partial\mathcal{S}$.

The results are summarized in \cref{fig:geodesic-distortion}, where we report for each pair of shapes the average distortion over the surface (in blue) and the volume (in orange), together with the standard deviation. We can see that, in all cases, the distortions are comparable, with the tetrahedral meshes generally achieving a slightly better preservation of the geodesic distances. As a consequence, we guess that, for most practical applications, the property of a pair of shapes of being isometric is preserved between a surface and volumetric representation, thus allowing us to safely extend the functional maps framework from surfaces to volumes.

\section{Boundary Trace of Eigenfunctions}\label{sec:boundarytrace}

\begin{figure}[t]
    \centering
    \includegraphics[width=\linewidth]{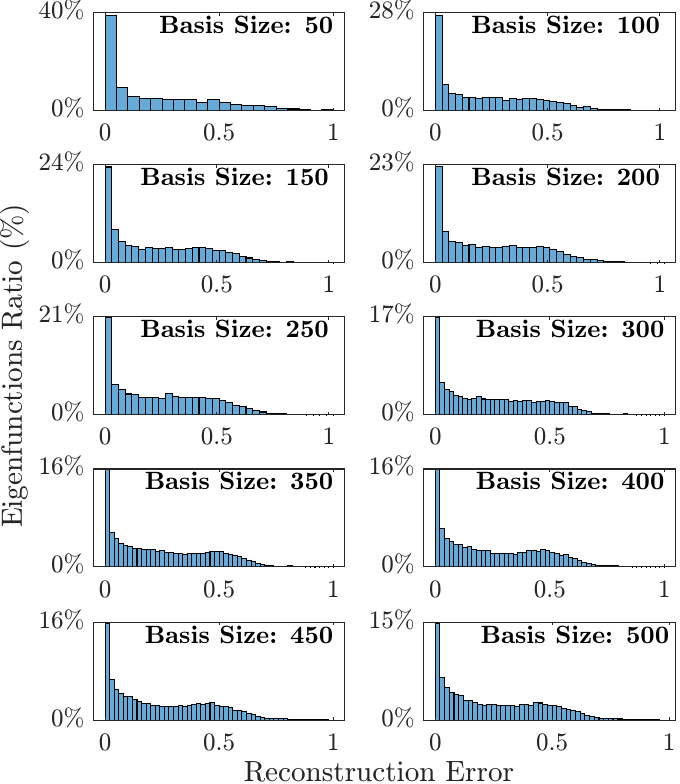}
    \caption{For a varying basis size, we try to reconstruct the LBO eigenfunctions of the surface using the restriction of the volumetric LBO eigenfunctions at the boundary. The reconstruction error lies in the range $[0, 1]$.}
    \label{fig:lbo-vol2surf}
    \vspace{-2em}
\end{figure}

\begin{figure*}[t]
	\centering    
	\includegraphics[width=\textwidth]{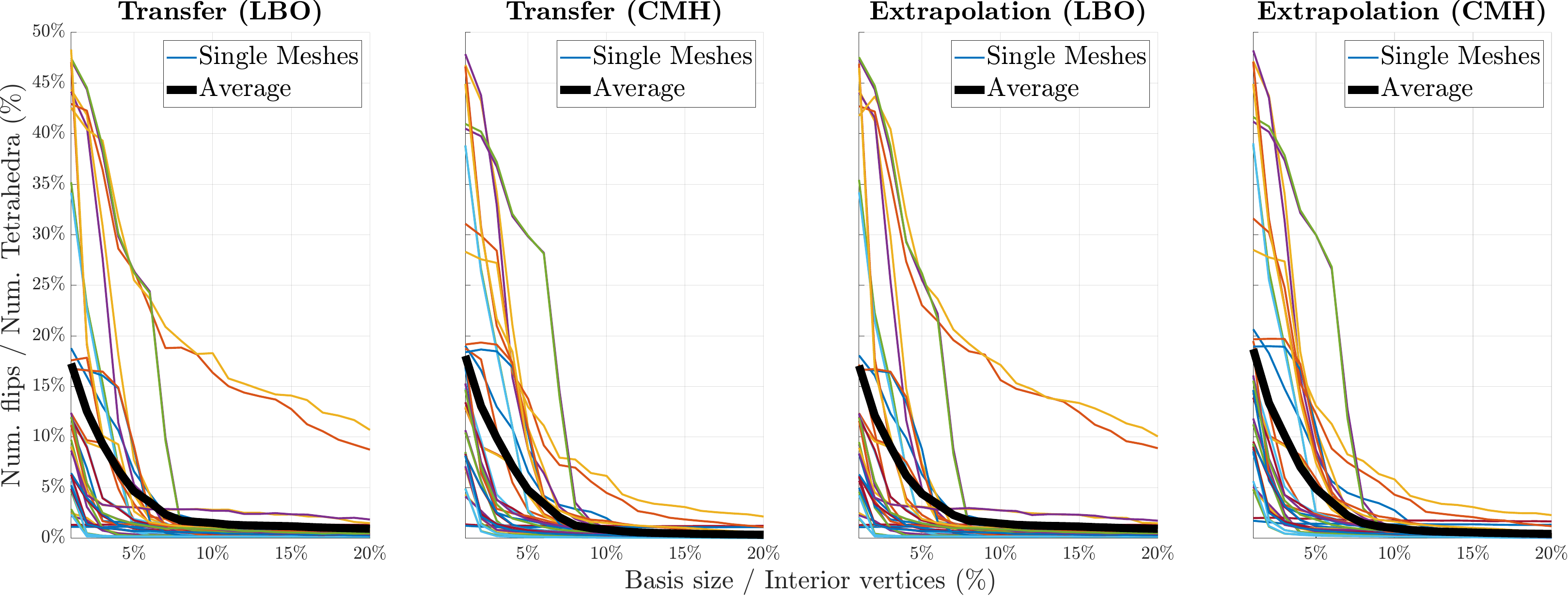}
	\caption{Number of flipped tetrahedra obtained by transferring the mesh connectivity considering a growing number of basis functions. From left to right: coordinate transfer using the LBO basis; coordinate transfer using the CMH basis; coordinate extrapolation using the LBO basis; coordinate extrapolation using the CMH basis. The smaller colored lines show the trend for each pair of shapes in our reference dataset~\cite{su2019practical}. The bold black line is the average trend across the entire dataset. The two pairs that take longer to converge with the LBO basis functions (yellow and orange lines) are the ones where the source and target differ the most and are strongly non-isometric. This explains why using only the intrinsic information from the LBO eigenbasis makes the method converge slowly and adding extrinsic information from CMH aligns the pairs with the other results.}
	\label{fig:connectivity-transfer}
\end{figure*}

As mentioned in \cref{sec:method:volmap}, given the volumetric LBO eigenfunctions $\evecs{\M}$ of a 3-dimensional manifold $\M$, the restriction $\evecs{\partial\M}$ of such basis at the boundary surface $\partial\M$ (often also called \emph{boundary trace}) forms a set of linearly independent functions over $\partial\M$~\cite{lasiecka1983stabilization,triggiani2008linear}.

However, $\evecs{\partial\M}$ is not necessarily equivalent to the LBO eigenbasis $\mathbf{\Psi}_{\partial\M}$ of the surface, meaning that it could be a sub-optimal basis~\cite{aflalo2015optimality,aflalo2016best}. In order to evaluate the expressive power of the boundary trace of the volumetric eigenfunctions, we try to reconstruct the basis $\mathbf{\Psi}_{\partial\M}$ using the basis $\evecs{\partial\M}$. For our experiment, we use the 80 shapes from the dataset presented by \citet{su2019practical} and we try to reconstruct the first $k$ eigenfunctions of the surface with the first $k$ boundary traces. In particular, we first compute the reconstruction $\mathbf{\Psi}' = \evecs{\partial\M}\evecs{\partial\M}\pinv\mathbf{\Psi}_{\partial\M}$. Then, for each surface eigenfunction $\psi_i$ and its reconstruction $\psi_i'$, we compute the reconstruction error as
\begin{equation}
    \mathrm{err}(\psi_i)
    =
    \int_{\partial\M} (\psi_i(x) - \psi_i'(x))^2\,\mathrm{d}x\,.
\end{equation}
This error measure lies in the range $[0, 1]$.

In \cref{fig:lbo-vol2surf} we show the distributions of the reconstruction errors across the entire dataset for different values of $k$. As shown in the figure, very few of the surface eigenfunctions are close to be orthogonal to the boundary traces. 
However, the boundary restriction of the volumetric eigenfunctions is able to faithfully represent a significant portion of the surface eigenbasis, and provides a reasonable approximation for most of them. Therefore, while not being the optimal basis, the boundary trace of the volumetric LBO eigenfunctions still provides a good candidate for representing a $k$-dimensional function space over the surface.

\section{Volumetric Functional Maps Metrics}\label{sec:volumemetrics}

\begin{table}[t]
    \centering
    \resizebox{\columnwidth}{!}{%
    \begin{tabular}{ccccc}
        \textbf{Method} & \textbf{Continuity} & \textbf{Coverage} & \textbf{Dirichlet} & \textbf{Flips} \\
        \hline
        FMaps &         1.32 & 91.48\% & 3.82\% & 10.03\%\\
        ZoomOut &       1.26 & 92.63\% & 3.80\% & 9.72\%\\
        ZoomOut Fast &  1.32 & 92.88\% & 3.79\% & 10.80\%\\
        Orthoprods &    1.29 & 87.78\% & 3.81\% & 5.80\%\\
    \end{tabular}
    }
    \caption{Additional metrics for the volumetric shape matching experiments from \cref{ssec:res_vol_map} on the dataset from \citet{su2019practical}.}
    \label{tab:volmap-extra-metrics}
\end{table}

Finally, we consider additional metrics to better evaluate the quality of volumetric shape matching discussed in \cref{ssec:res_vol_map}. In particular, we report in \cref{tab:volmap-extra-metrics} the continuity and coverage of the correspondence, as well as its Dirichlet energy, which provide a further characterization of maps in general~\cite[\S5]{ren:2018:continuousfmaps}. We also report the average percentage of flipped tetrahedra, which is an intrinsically volumetric measure. All the values are averaged across the entire dataset from \citet{su2019practical}. We stress that all reported values (including those in \cref{ssec:res_vol_map}) refer to volumetric estimates of the metrics. 

\section{Connectivity Transfer and Basis Size}\label{sec:volmapbsize}

While the average behavior of our connectivity transfer approaches is clear from the results shown in \cref{tab:connectivity-transfer}, there are additional information that can be better appreciated by examining the behavior of the methods on the single pairs. In particular, we can see from \cref{fig:connectivity-transfer} that some pairs are particularly challenging and results in a number of flips that is significantly higher than the average, when using the LBO eigenbasis. However, CMH seems to be a strong regularizer: the basis extended with the vertex coordinates does not seem to have a significant impact on the other curves, but it is definitely able to align the behavior of these outliers to the average. 

On the other hand, it seems that the introduction of the coordinate harmonics negatively affects smaller bases. Indeed, on the left end of the curves, the use of CMH as a basis seems to slow down the decrease rate of the curve.

In summary, by inspecting these results we educatedly guess that a good rule of thumb could be to use CMH when we have access to a larger basis, and stick to the LBO eigenbasis if we can only afford to compute a few basis functions.

We stress that using 20\% of the entire basis, in general, is computationally expensive, and we report the values only for analysis purposes. Indeed, for our experiments on the datasets from \citet{nigolian2023expansion} (see \cref{ssec:conn_trans}), we consider a significantly smaller basis with 500 eigenfunctions, covering from 0.5\% to 5\% of the total basis, depending on the resolution of the tetrahedral mesh.

\section{Volumetric Shape Matching Dataset}\label{sec:brokendata}

\begin{figure}[t]
	\centering
	\includegraphics[width=\linewidth]{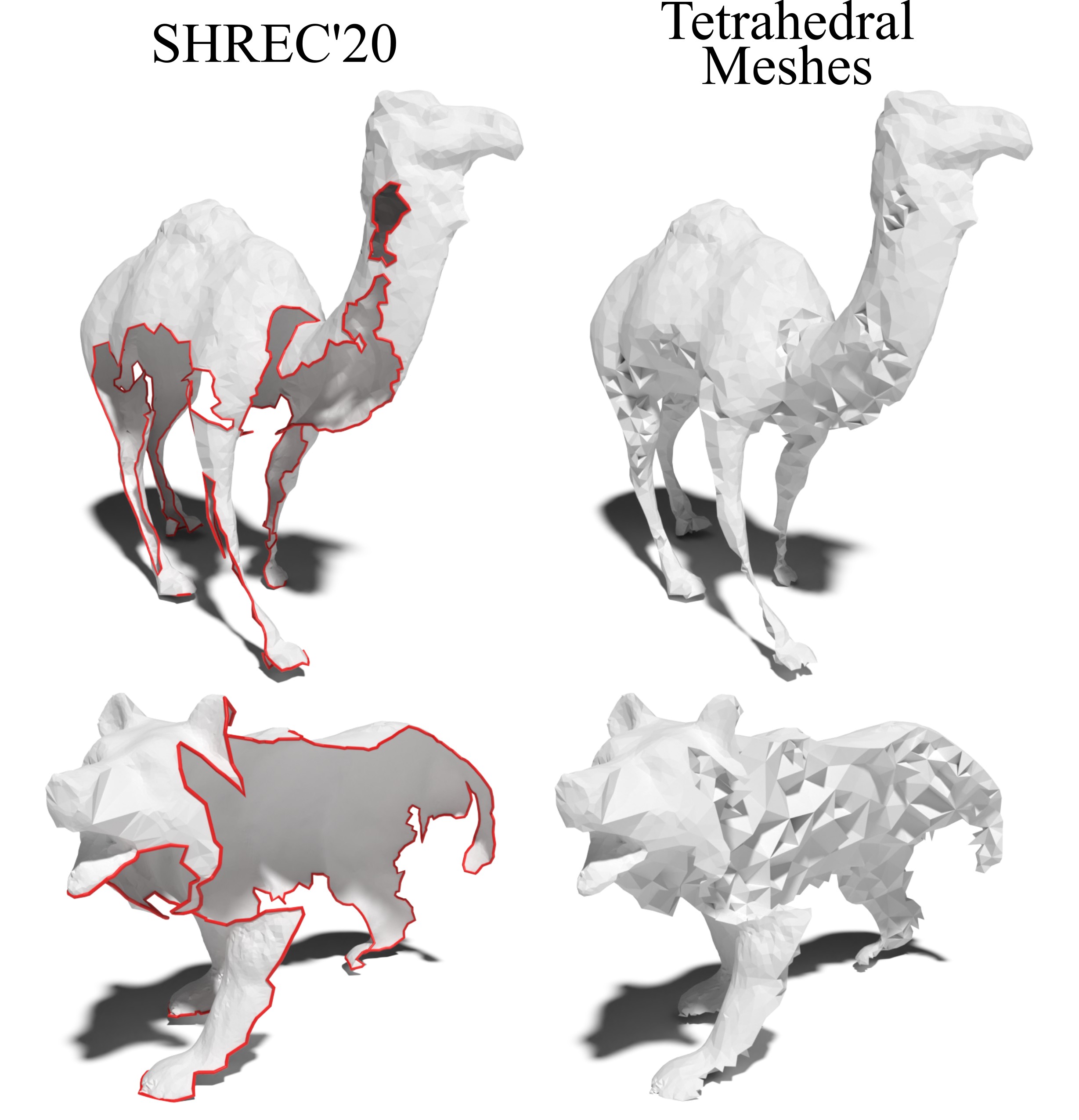}
	\caption{The two partial surfaces from the SHREC'20 dataset that we discarded from our experiments because once turned into tetrahedral meshes they do not correctly convey the intended shape, introducing an unwanted bias in the comparative analysis with surface-based methods.}
	\label{fig:shrec20_fail}
\end{figure}

As mentioned in \cref{ssec:res_matching}, well-established datasets for shape matching only provide surfaces. On top of that, most of the available recent datasets try to highlight some challenges in the problem of finding a correspondence; this includes partialities, holes, heavy topological noise, self-intersections, clutter, and so on.

Dropping the assumptions of manifoldness, watertightness, and absence of self-intersections for the surfaces, the most robust tool for producing tetrahedral meshes results to be fTetWild~\cite{hu2020fast}. Nonetheless, for some of the surfaces, the problem of finding a tetrahedralization is so ill-posed that any resulting volume would not provide meaningful information about the 3D shape.

Indeed, for SHREC'20, we discard two partial meshes (see~\cref{fig:shrec20_fail}) because the inside/outside system of fTetWild was not able to correctly reproduce a volumetric mesh that allows a meaningful comparison with surface methods. For the same reason, we are not able to consider alternative datasets such as SMAL~\cite{zuffi20173d}, which contains numerous self-intersections that yield topological changes during tetrahedralization. Attempts to sanitize these models with various combinations of geometry processing tools~\cite{portaneri2022alpha,hang2015tetgen,livesu2019cinolib,diazzi2021convex,diazzi2023constrained} ended with no luck. Indeed, our difficulties in creating a dataset that allows to meaningfully compare surface-based and volume-based shape matching methods raises an open research challenge that will hopefully be addressed by the community in the near future (\cref{sec:conclusions}).

\section{Choice of the Basis for Shape Matching}\label{sec:matchingbasis}

\begin{figure}[t]
    \centering
    \includegraphics[width=\columnwidth]{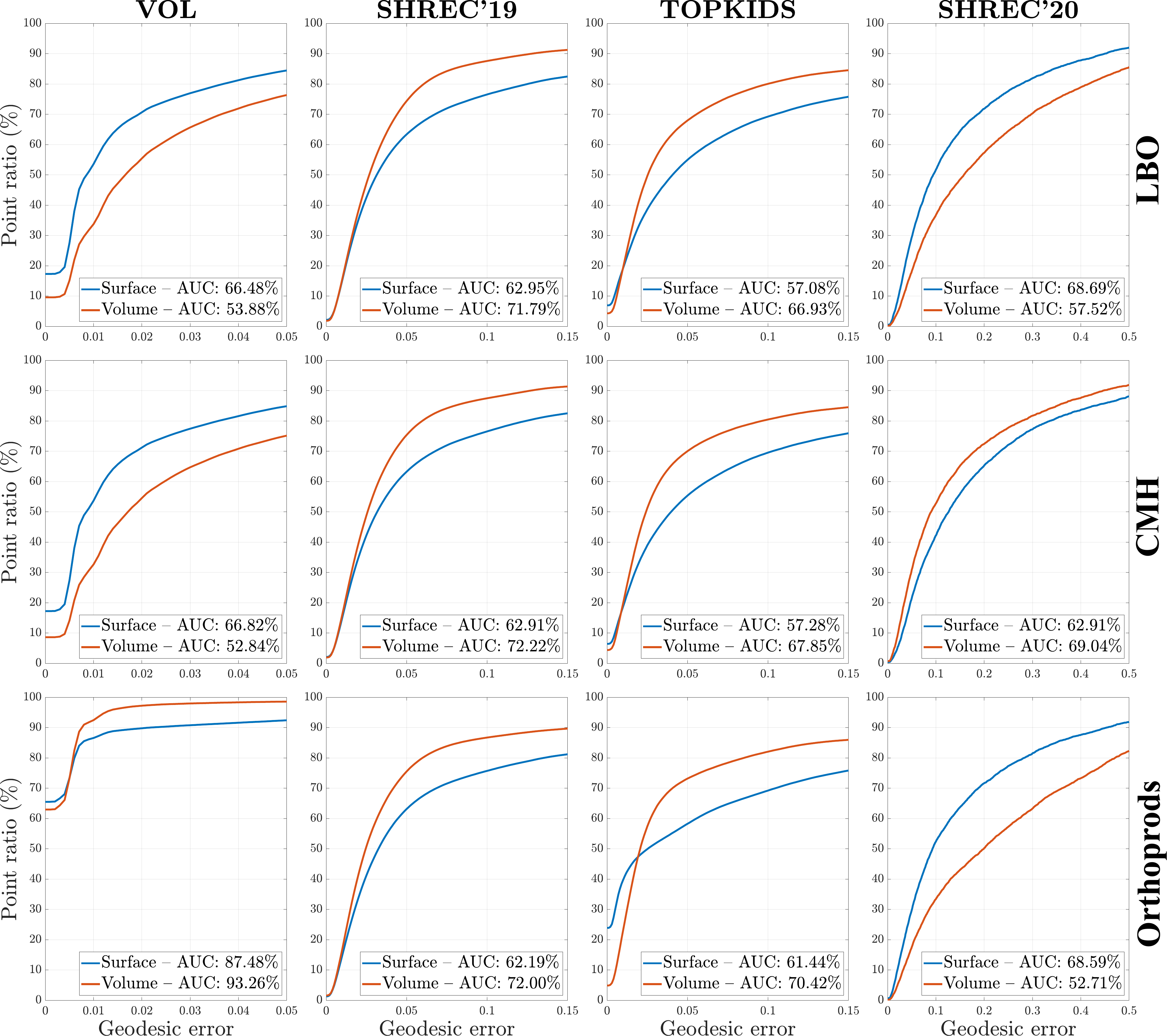}
    
    
    \caption{Cumulative plot of the geodesic error obtained with the standard functional maps framework (blue) on surfaces and our volumetric functional maps approach (orange) on four different datasets (columns) and using three different bases (rows). The results are averaged across each dataset.
    }
    \label{fig:surf-match}
\end{figure}

\begin{figure}[t]
    \centering
    \includegraphics[width=\columnwidth]{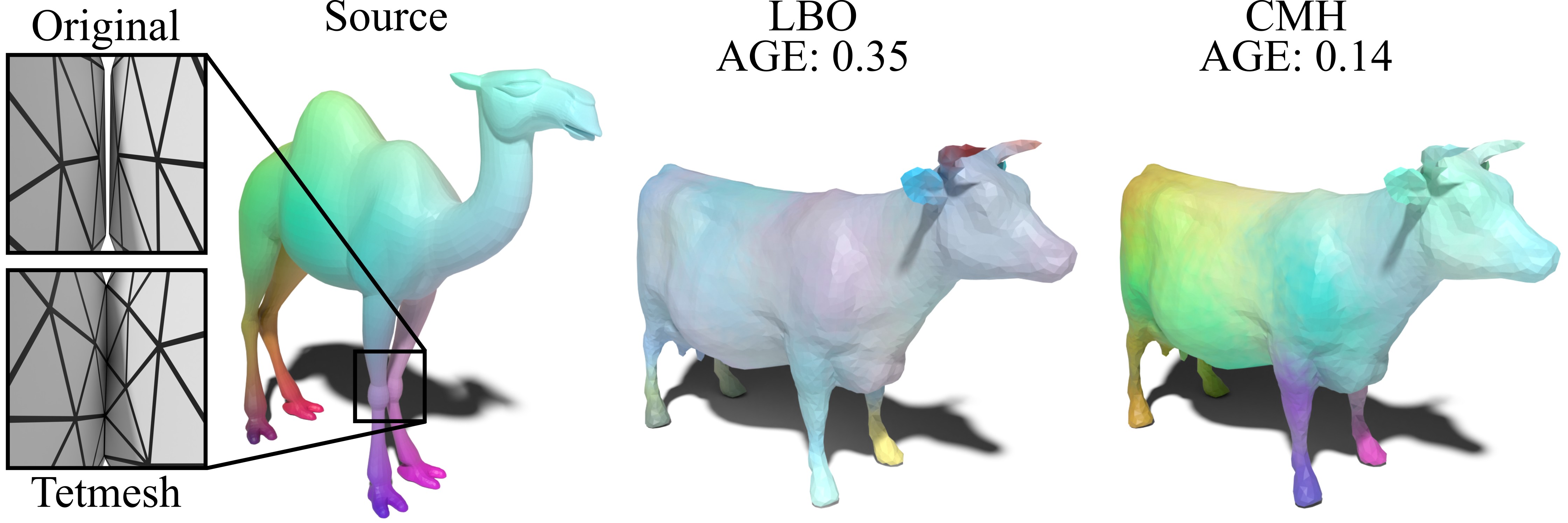}
    \caption{The tool we use to tetrahedralize surface matching datasets~\cite{hu2020fast} is robust, but occasionally introduces topological noise, closing unwanted handles that alter the mapping task (closeup). The LBO eigenbasis is not robust against such defects. Conversely, the CMH basis is robuster and greatly improves the mapping quality. Both the camel and the cow are from the SHREC'20 dataset.    
    }
    \label{fig:cmh-better}
\end{figure}

More detailed results about the comparison can be found in \cref{fig:surf-match}, where we show the cumulative geodesic error curves averaged across each dataset and their corresponding area under the curve (AUC). By comparing these curves against the results shown in \cref{tab:dataset-methods}, we can appreciate how the results summarized by the average geodesic error reflect on the cumulative curves, highlighting the consistently better behavior of our approach, as well as the importance in the choice of the basis.

In this regard, the use of the CMH basis seems to be particularly effective on SHREC'20, where the surface-based approach achieves better accuracy otherwise. Based on our understanding, this depends on the basis alignment algorithm we use~\cite{nogneng2017informative}, which is known to not be robust under strong non-isometric deformations and topological changes, which occasionally occurred when tetrahedralizing our surface-based reference datasets. This behavior can be partially compensated by using more informative bases such as CMH, which provide a better tool for representing extrinsic information. This can be better appreciated in the example from \cref{fig:cmh-better}, where we show a surface coordinate transfer in a pair from the SHREC'20 dataset with our volumetric approach using the standard LBO eigenbasis and the CMH basis. However, integrating more robust basis alignment methods is certainly worth of future investigation.

As a final note, for all our experiments we use the values suggested in the original papers when selecting the size of the bases. In particular, for the LBO eigenbasis we compute a $20\times20$ functional map using descriptors derived from 200 eigenfunctions for each shape. The map is extended up to the size $120\times120$ using ZoomOut, with a step size of 5 basis functions per iteration. In the case of CMH, the last 3 functions are replaced with the orthogonalized coordinates. For the Orthoprods basis, we use the suggested values of 40 eigenfunctions and second order polynomials. However, instead of the numerically unstable derivation of the map proposed in the original paper~\cite{maggioli2021orthogonalized}, we extend the functional map with ZoomOut.

\section{Algorithmic Details}\label{sec:algorithmics}

A sample MATLAB implementation of our volumetric functional maps framework is available at \url{https://github.com/filthynobleman/vol-fmaps}. Here we provide the pseudocode for both our connectivity transfer approaches (Sec. \ref{sec:method:volmap}) and our volume-aware surface correspondence method (Sec. \ref{sec:method:surfmap}).

Algorithm \ref{algo:conn-transfer} summarizes our solution for transfering connectivity through functional mapping. Firstly, we compute a tetrahedralization $\N$ for the interior of the target surface $\partial\N$ (Line 2). The volumetric meshes $\M$ and $\N$ are then put in functional correspondence via the map $\mathbf{C}$, which is approximated from the known alignment of the eigenfunctions at the surface (Lines 3-5). The source volume $\M$ is used to initialize the final mesh $\mathcal{R}$ (Line 6), whose coordinate are replaced from the functional transfer of the coordinates of $\N$.

\begin{algorithm}[t]
\caption{Functional connectivity transfer algorithm.}\label{algo:conn-transfer}
\begin{algorithmic}[1]
    \Procedure{Transfer}{$\M$, $\partial\N$, $\pi : \partial\M \to \partial\M$}
        \State $\N \gets$ computed interior of $\partial\N$
        \State $\evecs{\M} \gets \Call{LBODecomposition}{\M}$
        \State $\evecs{\N} \gets \Call{LBODecomposition}{\N}$
        \State $\mathbf{C} \gets \evecs{\partial\M}\pinv\,T_{\pi}(\evecs{\partial\N})$
        \State $\mathcal{R} \gets \M$
        \State $\myvec{x}_{\mathcal{R}} \gets \evecs{\M}\,\mathbf{C}\,\evecs{\N}\pinv\,\myvec{x}_{\N}$
        \State $\myvec{y}_{\mathcal{R}} \gets \evecs{\M}\,\mathbf{C}\,\evecs{\N}\pinv\,\myvec{y}_{\N}$
        \State $\myvec{z}_{\mathcal{R}} \gets \evecs{\M}\,\mathbf{C}\,\evecs{\N}\pinv\,\myvec{z}_{\N}$
        \State \Return $\mathcal{R}$
    \EndProcedure
\end{algorithmic}
\end{algorithm}

\begin{algorithm}[t]
\caption{Spectral coordinate extrapolation algorithm.}\label{algo:conn-extrapolate}
\begin{algorithmic}[1]
    \Procedure{Extrapolate}{$\M$, $\partial\N$, $\pi : \partial\M \to \partial\M$}
        \State $\evecs{\M} \gets \Call{LBODecomposition}{\M}$
        \State $\mathcal{R} \gets \M$
        \State $\myvec{x}_{\mathcal{R}} \gets \evecs{\M}\,\evecs{\partial\M}\pinv\,T_{\pi}(\myvec{x}_{\partial\N})$
        \State $\myvec{y}_{\mathcal{R}} \gets \evecs{\M}\,\evecs{\partial\M}\pinv\,T_{\pi}(\myvec{y}_{\partial\N})$
        \State $\myvec{z}_{\mathcal{R}} \gets \evecs{\M}\,\evecs{\partial\M}\pinv\,T_{\pi}(\myvec{z}_{\partial\N})$
        \State \Return $\mathcal{R}$
    \EndProcedure
\end{algorithmic}
\end{algorithm}

\begin{algorithm}[t]
\caption{Surface restricted volume matching algorithm.}\label{algo:vol2surf}
\begin{algorithmic}[1]
    \Procedure{Vol2Surf}{$\partial\M$, $\partial\N$}
        \State $\M \gets$ computed interior of $\partial\M$
        \State $\N \gets$ computed interior of $\partial\N$
        \State $\evecs{\M} \gets \Call{LBODecomposition}{\M}$
        \State $\evecs{\N} \gets \Call{LBODecomposition}{\N}$
        \State $\mathbf{C} \gets \Call{OptimizeVolFMAP}{\evecs{\M}, \evecs{\N}}$
        \State $\pi \gets \Call{NNSearch}{\evecs{\partial\M}\mathbf{C},\,\evecs{\partial\N}}$
        \State \Return $\pi$
    \EndProcedure
\end{algorithmic}
\end{algorithm}

The approach for extrapolating the interior volume from the surface is summarized in Algorithm \ref{algo:conn-transfer}. In this case, we only need the LBO eigenfunctions of $\M$ (Line 2). Again, we initialize a final mesh $\mathcal{R}$ from the source volume $\M$ (Line 3). However, with this approach the interior volume is obtained by projecting the coordinates at the surface in the functional space and reconstructing them in the whole volumetric mesh. The projection is performed using the boundary traces of the eigenfunctions, and the reconstruction is done with the full basis.

Finally, we summarize in Algorithm \ref{algo:vol2surf} our procedure to compute the surface correspondence by exploiting the volumetric information. We first compute the interior volumes $\M$ and $\N$ for the surfaces $\partial\M$ and $\partial\N$, respectively (Line 2-3). The LBO eigenbases are then computed on the entire volume (Lines 4-5) and put in functional correspondence through the optimization of the map $\mathbf{C}$ (Line 6). Finally, we compute the correspondence $\pi$ via nearest neighbor, limiting the search to the boundary traces of the eigenfunctions.


\end{document}